\documentclass{elsart}

\usepackage{times}
\usepackage{helvetic}

\usepackage{amssymb}

\usepackage{graphicx}

\def\portlandEUniLinks{20\,024}

\def\portlandAllUniLinks{200\,000}

\begin{document}

\begin{frontmatter}
\title{Parallel implementation of the TRANSIMS micro-simulation}

\author[eth]{Kai Nagel\thanksref{kn}}
\author[sdm]{and Marcus Rickert\thanksref{mr}}

\address[eth]{Dept.\ of Computer Science, ETH Z{\"u}rich, Switzerland}
\address[sdm]{sd\&m AG, Troisdorf, Germany}

\thanks[kn]{nagel@inf.ethz.ch.  Postal: ETH Zentrum IFW B27.1,
  8092~Z{\"u}rich, Switzerland}
\thanks[mr]{marcus.rickert@topmail.de}

\maketitle

\begin{abstract}\noindent
  This paper describes the parallel implementation of the TRANSIMS
  traffic micro-simulation.  The parallelization method is domain
  decomposition, which means that each CPU of the parallel computer is
  responsible for a different geographical area of the simulated
  region.  We describe how information between domains is exchanged,
  and how the transportation network graph is partitioned.  An
  adaptive scheme is used to optimize load balancing.  
  
  We then demonstrate how computing speeds of our parallel
  micro-si\-mu\-lations can be systematically predicted once the
  scenario and the computer architecture are known.  This makes it
  possible, for example, to decide if a certain study is feasible with
  a certain computing budget, and how to invest that budget.  The main
  ingredients of the prediction are knowledge about the parallel
  implementation of the micro-simulation, knowledge about the
  characteristics of the partitioning of the transportation network
  graph, and knowledge about the interaction of these quantities with
  the computer system.  In particular, we investigate the differences
  between switched and non-switched topologies, and the effects of
  10~Mbit, 100~Mbit, and Gbit Ethernet.
  
  As one example, we show that with a common technology -- 100~Mbit
  switched Ethernet -- one can run the 20\,000-link EMME/2-network for
  Portland (Oregon) more than 20 times faster than real time on
  16~coupled Pentium~CPUs.
\end{abstract}

\begin{keyword}
Traffic simulation, parallel computing, transportation planning, TRANSIMS
\end{keyword}

\end{frontmatter}

\section{Introduction}

It is by now widely accepted that it is worth investigating if the
microscopic simulation of large transportation
systems~\cite{PARAMICS:Supercomp,TRANSIMS} is a useful addition to the existing
set of tools.  By ``microscopic'' we mean that all entities of the
system -- travelers, vehicles, traffic lights, intersections, etc. --
are represented as individual objects in the
simulation~\cite{NETSIM,NETSIM:new,AIMSUN,INTEGRATION,DYNAMIT:MITSIM,HUTSIM,VISSIM}.

The conceptual advantage of a micro-simulation is that in principle it
can be made arbitrarily realistic.  Indeed, microscopic simulations
have been used for many decades for problems of relatively small
scale, such as intersection design or signal phasing.  What is new is
that it is now possible to use microscopic simulations also for really
large systems, such as whole regions with several millions of
travelers.  At the heart of this are several converging
developments:\begin{enumerate}

\item
The advent of fast desktop workstations.   

\item The possibility to connect many of these workstations to
  parallel supercomputers, thus multiplying the available computing
  power.  This is particularly attractive for agent-based
  transportation simulations since they do not benefit from
  traditional vector supercomputers.
  
\item In our view, there is a third observation that is paramount to
  make these approaches work: many aspects of a ``correct''
  macroscopic behavior can be obtained with rather simple microscopic
  rules.  

\end{enumerate}
The third point can actually be rigorously proven for some cases.  For
example, in physics the ideal gas equation, $p V = m R T$, can be
derived from particles without any interaction, i.e.\ they move
\emph{through} each other.  For traffic, one can show that rather
simple microscopic models generate certain fluid-dynamical equations
for traffic flow~\cite{Nagel:flow:trb}.
  
In consequence, for situations where one expects that the
fluid-dynamical representation of traffic is realistic enough for the
dynamics but one wants access to individual vehicles/drivers/..., a
simple microscopic simulation may be the solution.  In addition to
this, with the microscopic approach it is always possible to make it
more realistic at some later point.  This is much harder and sometimes
impossible with macroscopic models.


The TRANSIMS (TRansportation ANalysis and SIMulation System) project
at Los Alamos National Laboratory~\cite{TRANSIMS} is such a
micro-simulation project, with the goal to use micro-simulation for
transportation planning.  Transportation planning is typically done
for large regional areas with several millions of travelers, and it
is done with 20~year time horizons.  The first means that, if we want
to do a micro-simulation approach, we need to be able to simulate
large enough areas fast enough.  The second means that the methodology
needs to be able to pick up aspects like induced travel, where people
change their activities and maybe their home locations because of
changed impedances of the transportation system.  As an answer,
TRANSIMS consists of the following modules:\begin{itemize}
  
\item \textbf{Population generation}.  Demographic data is
  disaggregated so that we obtain individual households and individual 
  household members, with certain characteristics, such as a street
  address, car ownership, or household income~\cite{Beckman:pops}.
  
\item \textbf{Activities generation}.  For each individual, a set of
  activities and activity locations for a day is
  generated~\cite{Vaughn:etc:HATPs,Bowman:thesis}.
  
\item \textbf{Modal and route choice}.  For each individual, modes and
  routes are generated that connect activities at different
  locations~\cite{Jacob:etc:comp}.
  
\item \textbf{Traffic micro-simulation}. Up to here, all individuals
  have made \emph{plans} about their behavior.  The traffic
  micro-simulation executes all those plans simultaneously.  In
  particular, we now obtain the result of \emph{interactions} between
  the plans -- for example congestion.\footnote{%
    It is sometimes argued that TRANSIMS is unnecessarily realistic
    for the questions it is supposed to answer.  Although we tend to
    share the same intuition (see, for example, our work on the
    so-called queue model~\protect\cite{Simon:Nagel:queue}), we think
    that this needs to be evaluated systematically.  We also expect
    that the answer will depend on the precise question: It will be
    possible to answer certain questions with very simple models,
    while other questions may need much more realistic models.
}

\end{itemize}
As is well known, such an approach needs to make the modules
consistent with each other: For example, plans depend on congestion,
but congestion depends on plans.  A widely accepted method to resolve
this is systematic relaxation~\cite{DYNAMIT:MITSIM} -- that is, make
preliminary plans, run the traffic micro-simulation, adapt the plans,
run the traffic micro-simulation again, etc., until consistency
between modules is reached.  The method is somewhat similar to the
Frank-Wolfe-algorithm in static assignment.

The reason why this is important in the context of this paper is that
it means that the micro-simulation needs to be run more than once --
in our experience about fifty times for a relaxation from
scratch~\cite{Rickert:thesis,Rickert:Nagel:isttt}.  In consequence, a
computing time that may be acceptable for a single run is no longer
acceptable for such a relaxation series -- thus putting an even higher
demand on the technology.

This can be made more concrete by the following
arguments:\begin{itemize}
  
\item The number of ``about fifty'' iterations was gained from
  systematic computational experiments using a scenario in Dallas/Fort
  Worth.  In fact, for route assignment alone, about twenty iterations
  are probably sufficient~\cite{Rickert:thesis,Rickert:Nagel:isttt},
  but if one also allows for other behavioral changes, more iterations
  are needed~\cite{Esser:Nagel:fvu}.  The numbers become plausible via
  the following argument: Since relaxation methods rely on the fact
  that the situation does not change too much from one iteration to
  the next, changes have to be small.  Empirically, changing more than
  10\% of the travellers sometimes leads to strong fluctuations away
  from relaxation~\cite{Rickert:thesis,Rickert:Nagel:isttt}.  A
  replanning fraction of 10\% means that we need 10~iterations in
  order to replan each traveller exactly once; and since during the
  first couple of iterations travellers react to non-relaxed traffic
  patterns, we will have to replan those a second time, resulting in
  15-20~iterations.  Nevertheless, future research will probably find
  methods to decrease the number of iterations.
  
\item We assume that results of a scenario run should be available
  within a few days, say two. Otherwise research becomes frustratingly
  slow, and we would assume that the same is true in practical
  applications.  Assuming further that we are interested in 24~hour
  scenarios, and disregarding computing time for other modules besides
  the microsimulation, this means that the simulation needs to run
  25~times faster than real time.

\end{itemize}
We will show in this paper that the TRANSIMS microsimulation indeed
can be run with this computational speed, and that, for certain
situations, this can even be done on relatively modest hardware.  By
``modest'' we mean a cluster of 10-20 standard PCs connected via
standard LAN technology (Beowulf cluster).  We find that such a
machine is affordable for most university engineering departments, and
we also learn from people working in the commercial sector (mostly
outside transportation) that this is not a problem.  In consequence,
TRANSIMS can be used without access to a supercomputer.  As mentioned
before, it is beyond the scope of this paper to discuss for which
problems a simulation as detailed as TRANSIMS is really necessary and
for which problems a simpler approach might be sufficient.

This paper will concentrate on the microsimulation of TRANSIMS.  The
other modules are important, but they are less critical for computing
(see also Sec.~\ref{sec:other}).  We start with a description of the
most important aspects of the TRANSIMS driving logic
(Sec.~\ref{sec:driving}).  The driving logic is designed in a way that
it allows domain decomposition as a parallelization strategy, which is
explained in Sec.~\ref{parallelization}.
We then demonstrate that the \emph{implemented} driving logic
generates realistic macroscopic traffic flow.  Once one knows that the
microsimulation can be partitioned, the question becomes how to
partition the street network graph.  This is described in
Sec.~\ref{sec:partitioning}.  Sec.~\ref{load_feedback} discusses how
we adapt the graph partitioning to the different computational loads
caused by different traffic on different streets.  These and
additional arguments are then used to develop a methodology for the
prediction of computing speeds (Sec.~\ref{sec:predict}).  This is
rather important, since with this one can predict if certain
investments in one's computer system will make it possible to run
certain problems or not.  We then shortly discuss what all this means
for complete studies (Sec.~\ref{sec:other}).  This is followed by a
summary.

\section{Related work}

As mentioned above, micro-simulation of traffic, that is, the
individual simulation of each vehicle, has been done for quite some
time (e.g.~\cite{Gerlough:CA}).  A prominent example is
NETSIM~\cite{NETSIM,NETSIM:new}, which was developed in the 70s.
Newer models are, e.g., the Wiedemann-model~\cite{Wiedemann:model},
AIMSUN~\cite{AIMSUN}, INTEGRATION~\cite{INTEGRATION},
MITSIM~\cite{DYNAMIT:MITSIM}, HUTSIM~\cite{HUTSIM}, or
VISSIM~\cite{VISSIM}.

NETSIM was even tried on a vector
supercomputer~\cite{Mahmassani:NETSIM:HPC}, without a real
break-through in computing speeds.  But, as pointed out earlier,
ultimately the inherent structure of agent-based micro-simulation is
at odds with the computer architecture of vector supercomputers, and
so not much progress was made on the supercomputing end of
micro-simulations until the parallel supercomputers became available.
One should note that the programming model behind so-called Single
Instruction Multiple Data (SIMD) parallel computers is very similar to
the one of vector supercomputers and thus also problematic for
agent-based simulations.  In this paper, when we talk about parallel
computers, we mean in all cases Multiple Instruction Multiple Data
(MIMD) machines.

Early use of parallel computing in the transportation community
includes parallelization of fluid-dynamical models for
traffic~\cite{parallel:fdyn4traff} and parallelization of assignment
models~\cite{Hislop:CONTRAM:parallel}.  Early implementations of
parallel micro-simulations can be found
in~\cite{Chang:etc:early:parallel:micro,THOREAU:performance,BaNaRi:1}.

It is usually easier to make an efficient parallel implementation from
scratch than to port existing codes to a parallel computer.  Maybe for
that reason, early traffic agent-based traffic micro-simulations which
used parallel computers were completely new designs and
implementations~\cite{PARAMICS:Supercomp,TRANSIMS,BaNaRi:1,THOREAU:performance}.
All of these use \emph{domain decomposition} as their parallelization
strategy, which means that the partition the network graph into
domains of approximately equal size, and then each CPU of the parallel
computer is responsible for one of these domains.  It is maybe no
surprise that the first three use, at least in their initial
implementation, some cellular structure of their road representation,
since this simplifies domain decomposition, as will be seen later.
Besides the large body of work in the physics community
(e.g.~\cite{Wolfram:book}), such ``cellular'' models also have some
tradition in the transportation
community~\cite{Gerlough:CA,Cremer:Ludwig}.

Note that domain decomposition is rather different from a functional
parallel decomposition, as for example done by
DYNAMIT/MITSIM~\cite{DYNAMIT:MITSIM}.  A functional decomposition means that
different modules can run on different computers.  For example, the
micro-simulation could run on one computer, while an on-line routing
module could run on another computer.  While the functional
decomposition is somewhat easier to implement and also is less demanding on
the hardware to be efficient, it also poses a severe limitation on the
achievable speed-up.  With functional decomposition, the maximally
achievable speed-up is the number of functional modules one can
compute simultaneously -- for example micro-simulation, router, demand
generation, ITS logic computation, etc.  Under normal circumstances,
one probably does not have more than a handful of these functional
modules that can truly benefit from parallel execution, restricting
the speed-up to five or less.  In contrast, as we will see the domain
decomposition can, on certain hardware, achieve a more than 100-fold
increase in computational speed.

In the meantime, some of the ``pre-existing'' micro-simulations are
ported to parallel computers.  For example, this has recently been
done for AIMSUN2~\cite{AIMSUN:parallel} and for DYNEMO~\cite{DYNEMO,DYNEMO:parallel},\footnote{%
  DYNEMO is not strictly a micro-simulation -- it has individual
  travelers but uses a macroscopic approach for the speed calculation.
  It is mentioned here because of the parallelization effort.
  } and a parallelization is planned for
VISSIM~\cite{VISSIM} (M.~Fellendorf, personal communication).

\section{Microsimulation driving logic}
\label{sec:driving}

The TRANSIMS-1999\footnote{%
  There are two versions of TRANSIMS with the number ``1.0'': One from
  1997, ``TRANSIMS Release 1.0''~\cite{Beckman:etc:case-study}, which
  we will refer to as ``TRANSIMS-1997'', and one from 1999,
  ``TRANSIMS--LANL--1.0''~\cite{TRANSIMS:rel1999}, which we will refer
  to as ``TRANSIMS-1999''.  From 1997 to 1999, many features were
  added, such as public transit with a different driving logic, or the
  option of using continuous corrections to the cellular structure.
  For the purposes of this paper, the differences are not too
  important, except that computational performance was also
  considerably improved.
} microsimulation uses a cellular automata (CA)
technique for representing driving dynamics
(e.g.~\cite{Nagel:flow:trb}).  The road is divided into cells, each of
a length that a car uses up in a jam -- we currently use 7.5~meters.
A cell is either empty, or occupied by exactly one car.  Movement
takes place by \emph{hopping} from one cell to another; different
vehicle speeds are represented by different hopping distances.  Using
one second as the time step works well (because of reaction-time
arguments~\cite{Krauss:thesis}); this implies for example that a
hopping speed of 5~cells per time step corresponds to 135~km/h.  This
models ``car following''; the rules for car following in the CA are:
(i)~linear acceleration up to maximum speed if no car is ahead;
(ii)~if a car is ahead, then adjust velocity so that it is
proportional to the distance between the cars (constant time headway);
(iii)~sometimes be randomly slower than what would result from (i) and
(ii).

Lane changing is done as pure sideways movement in a sub-time-step
before the forwards movement of the vehicles, i.e.\ each time-step is
subdivided into two sub-time-steps. The first sub-time-step is used
for lane changing, while the second sub-time-step is used for forward
motion.  Lane-changing rules for TRANSIMS are symmetric and consist of
two simple elements: Decide that you want to change lanes, and check
if there is enough gap to ``get
in''~\cite{Rickert:etc:2lane}.  A ``reason to change
lanes'' is either that the other lane is faster, or that the driver
wants to make a turn at the end of the link and needs to get into the
correct lane.  In the latter case, the accepted gap decreases with
decreasing distance to the intersection, that is, the driver becomes
more and more desperate.

Two other important elements of traffic simulations are signalized
turns and unprotected turns.  The first of those is modeled by
essentially putting a ``virtual'' vehicle of maximum velocity zero at
the end of the lane when the traffic light is red, and to remove it
when it is green.  Unprotected turns get modeled via ``gap
acceptance'': There needs to be a large enough gap on the priority
street for the car from the non-priority street to accept
it~\cite{HCM:94}.

A full description of the TRANSIMS driving logic would go beyond the
scope of the present paper.  It can be found in
Refs.~\cite{Nagel:etc:flow-char,TRANSIMS:rel1999}.

\section{Micro-simulation parallelization: Domain decomposition}
\label{parallelization}

An important advantage of the CA is that it helps with the design of a
parallel and local simulation update, that is, the state at time step
$t+1$ depends only on information from time step $t$, and only from
neighboring cells.  (To be completely correct, one would have to
consider our sub-time-steps.)  This means that domain decomposition
for parallelization is straightforward, since one can communicate the
boundaries for time step $t$, then locally on each CPU perform the
update from $t$ to $t+1$, and then exchange boundary information
again.

\begin{figure}[htb]
  \begin{center}
    \includegraphics[width=\hsize]{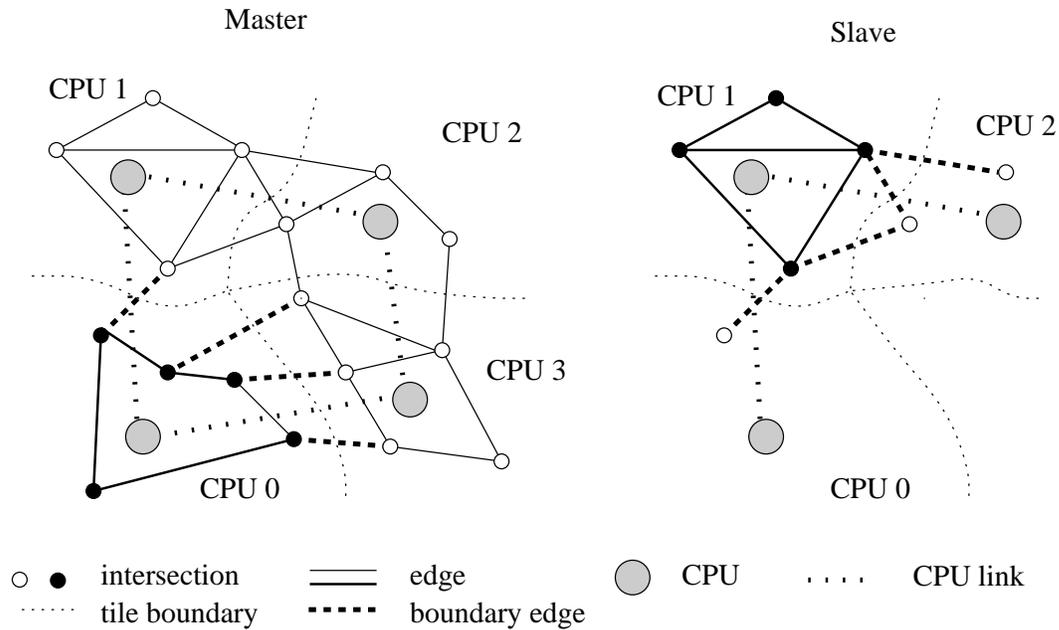}
    \caption{Domain decomposition of transportation network.  \emph{Left:} Global view.  \emph{Right:} View of a slave CPU.  The slave CPU is only aware of the part of the network which is attached to its local nodes.  This includes links which are shared with neighbor domains.}
    \label{fig:distrib}
  \end{center}
\end{figure}

\begin{figure}[htb]
  \begin{center}
    \includegraphics[width=\hsize]{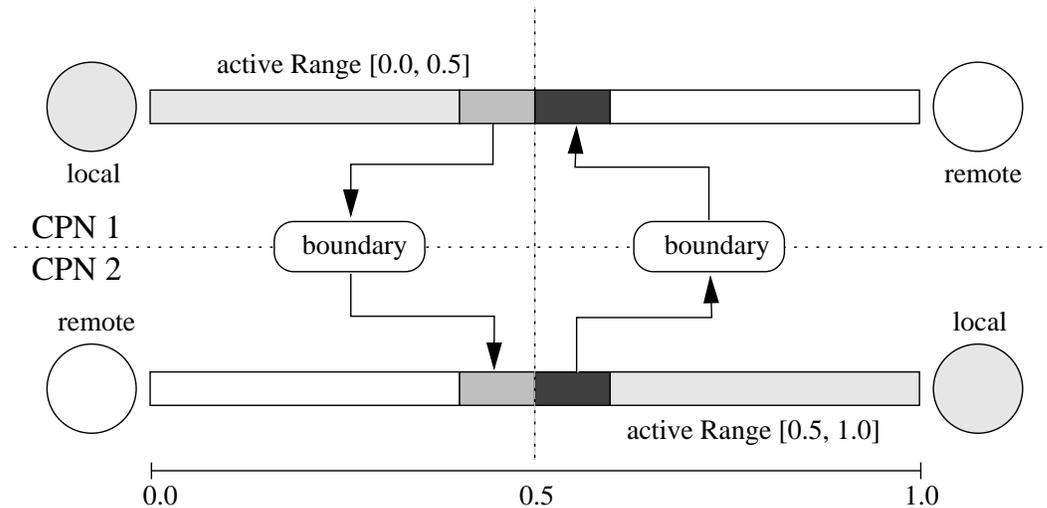}
    \caption{Distributed link.}
    \label{fig:bnd}
  \end{center}
\end{figure}

Domain decomposition means that the geographical region is decomposed
into several domains of similar size (Fig.~\ref{fig:distrib}), and
each CPU of the parallel computer computes the simulation dynamics for
one of these domains.  Traffic simulations fulfill two conditions
which make this approach efficient:
\begin{itemize}
  
\item Domains of similar size: The street network can be partitioned
  into domains of similar size. A realistic measure for size is the
  accumulated length of all streets associated with a domain.
  
\item Short-range interactions: For driving decisions, the distance of
  interactions between drivers is limited.  In our CA implementation,
  on links all of the TRANSIMS-1999 rule sets have an interaction
  range of $37.5$~meters ($=$ 5~cells) which is small with respect to
  the average link length.  Therefore, the network easily decomposes
  into independent components.

\end{itemize}
We decided to cut the street network in the middle of links rather
than at intersections (Fig.~\ref{fig:bnd}); THOREAU does the
same~\cite{THOREAU:performance}.  This separates the traffic
complexity at the intersections from the complexity caused by the
parallelization and makes optimization of computational speed easier.

In the implementation, each divided link is fully represented in both
CPUs.  Each CPU is responsible for one half of the link.  In order to
maintain consistency between CPUs, the CPUs send information about the
first five cells of ``their'' half of the link to the other CPU.  Five
cells is the interaction range of all CA driving rules on a link.  By
doing this, the other CPU knows enough about what is happening on the
other half of the link in order to compute consistent traffic.

\begin{figure}[htb]
  \begin{center}
    \includegraphics[height=0.8\textheight]{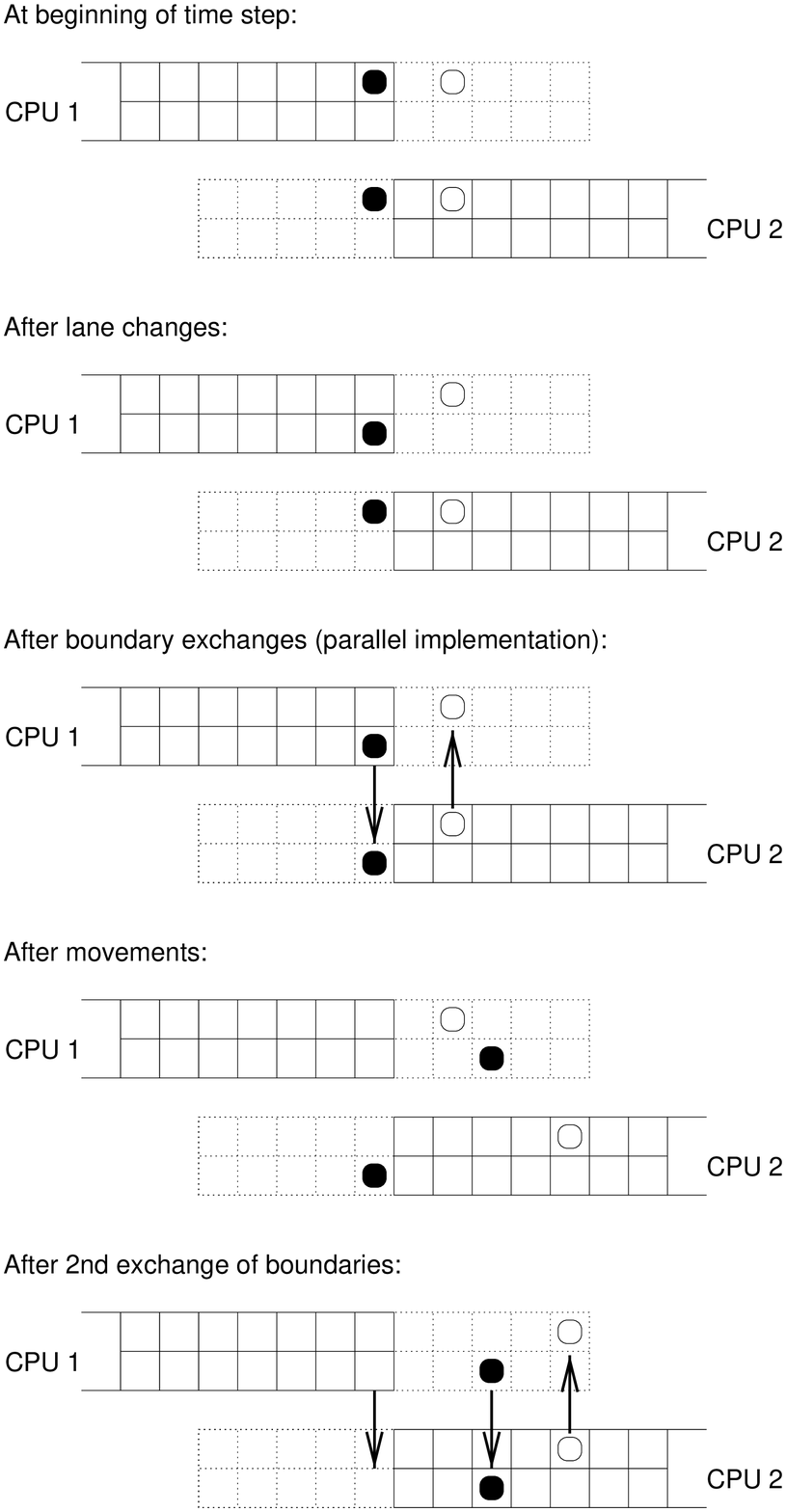}
    \caption{%
      Example of parallel logic of a split link with two lanes.
      The figure shows the general logic of one time step.  Remember
      that with a split link, one CPU is responsible for one
      half and another CPU is responsible for the other half.  These
      two halves are shown separately but correctly lined up.  The
      dotted part is the ``boundary region'', which is where the link
      stores information from the other CPU.  The arrows denote when
      information is transferred from one CPU to the other via
      boundary exchange.
}
    \label{fig:parallel}
  \end{center}
\end{figure}

The resulting simplified update sequence on the split links is
as follows (Fig.~\ref{fig:parallel}):\footnote{%
  Instead of ``split links'', the terms ``boundary links'', ``shared
  links'', or ``distributed links'' are sometimes used.  As is well
  known, some people use ``edge'' instead of ``link''.
}\begin{itemize}

\item Change lanes.

\item Exchange boundary information.

\item Calculate speed and move vehicles forward.

\item Exchange boundary information.

\end{itemize}
The TRANSIMS-1999 microsimulation also includes vehicles that enter
the simulation from parking and exit the simulation to parking, and
logic for public transit such as buses.  These additions are
implemented in a way that no further exchange of boundary information
is necessary.  

The implementation uses the so-called master-slave approach.
Master-slave approach means that the simulation is started up by a
master, which spawns slaves, distributes the workload to them, and
keeps control of the general scheduling.  Master-slave approaches
often do not scale well with increasing numbers of CPUs since the
workload of the master remains the same or even increases with
increasing numbers of CPUs.  For that reason, in TRANSIMS-1999 the
master has nearly no tasks except initialization and synchronization.
Even the output to file is done in a decentralized fashion.  With the
numbers of CPUs that we have tested in practice, we have never
observed the master being the bottleneck of the parallelization.

The actual implementation was done by defining descendent C++ classes
of the C++ base classes provided in a Parallel Toolbox. The underlying
communication library has interfaces for both PVM (Parallel Virtual
Machine~\cite{PVM}) and MPI (Message Passing Interface~\cite{MPI}).
The toolbox implementation is not specific to transportation
simulations and thus beyond the scope of this paper.  More information
can be found in \cite{Rickert:thesis}.

\section{Macroscopic (emergent) traffic flow characteristics}

In our view, it is as least as important to discuss the resulting
traffic flow characteristics as to discuss the details of the driving
logic.  For that reason, we have performed systematic validation of
the various aspects of the emerging flow behavior.  Since the
microsimulation is composed of car-following, lane changing,
unprotected turns, and protected turns, we have corresponding
validations for those four aspects.  Although we claim that this is a
fairly systematic approach to the situation, we do not claim that our
validation suite is complete.  For example,
weaving~\cite{INTEGRATION:weaving} is an important candidate for
validation.

It should be noted that we do not only validate our driving logic, but
we validate the \emph{implementation} of it, including the parallel
aspects.  It is easy to add unrealistic aspects in a parallel
implementation of an otherwise flawless driving logic; and the authors
of this paper are sceptic about the feasibility of formal verification
procedures for large-scale simulation software.

\begin{figure}[htb]
\halign to\hsize{&\hfill # \hfill\cr
    \includegraphics[width=0.49\hsize]{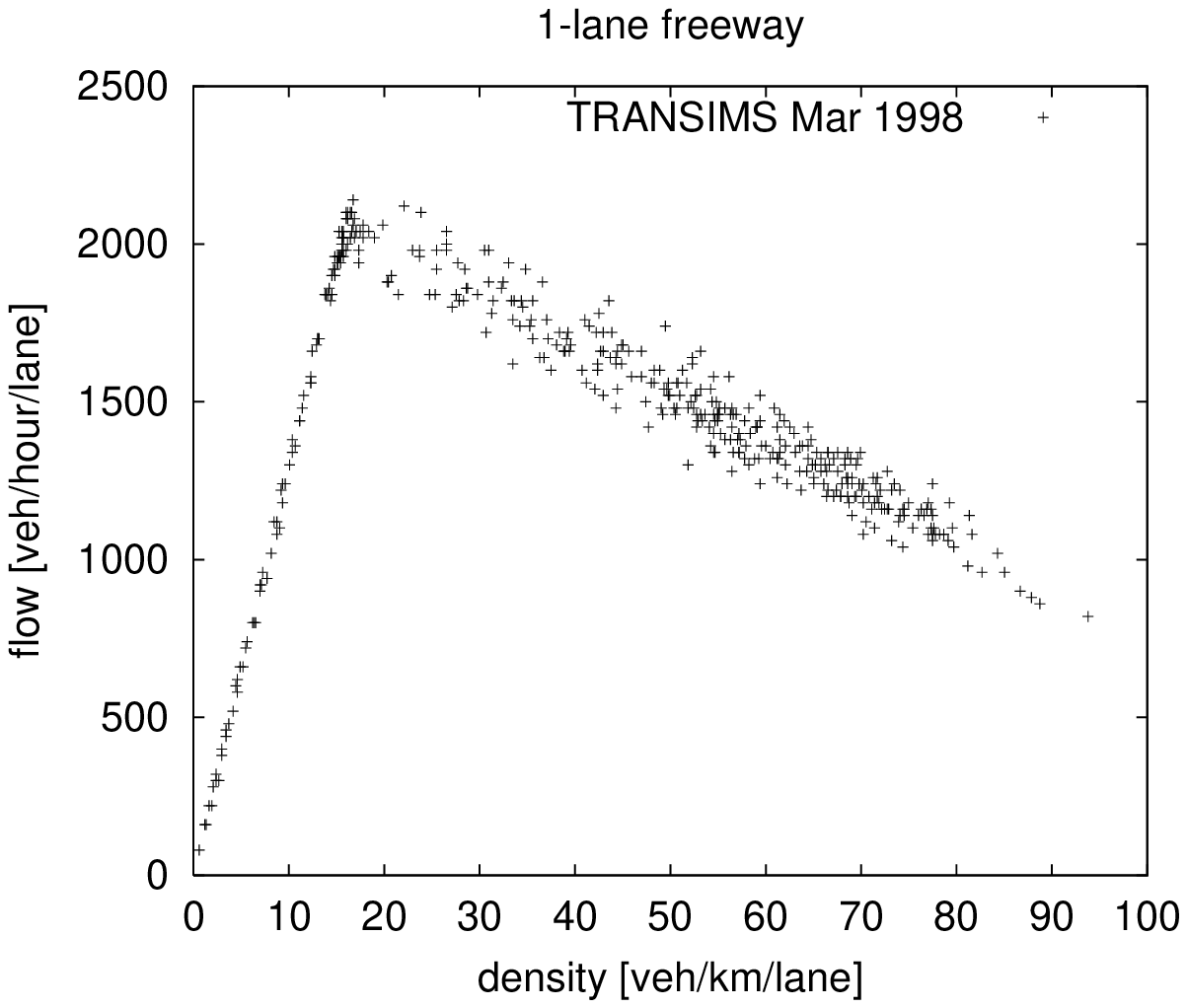}
&
    \includegraphics[width=0.49\hsize]{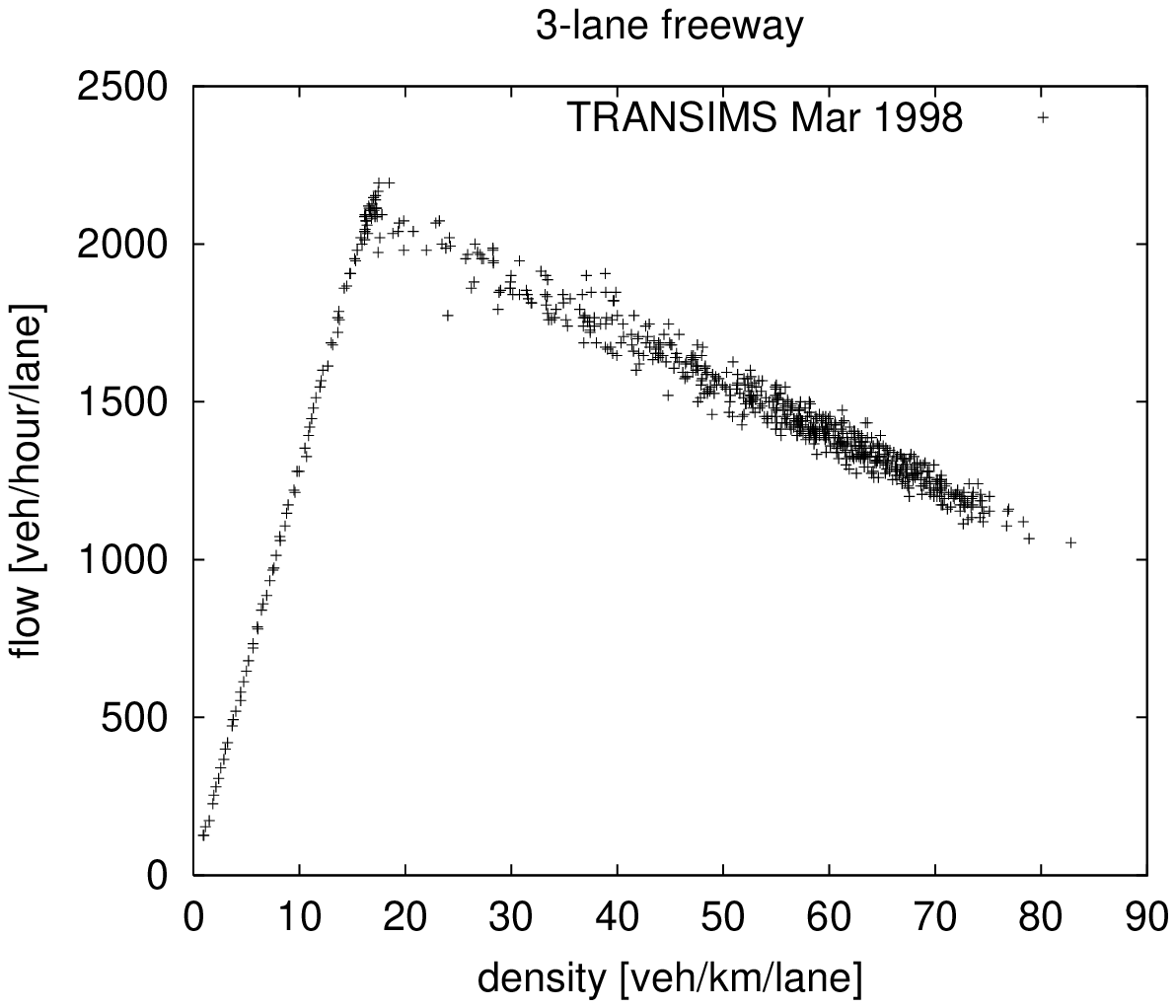}
\cr
}
\centerline{%
\parbox[c]{0.49\textwidth}{%
\includegraphics[width=0.49\textwidth]{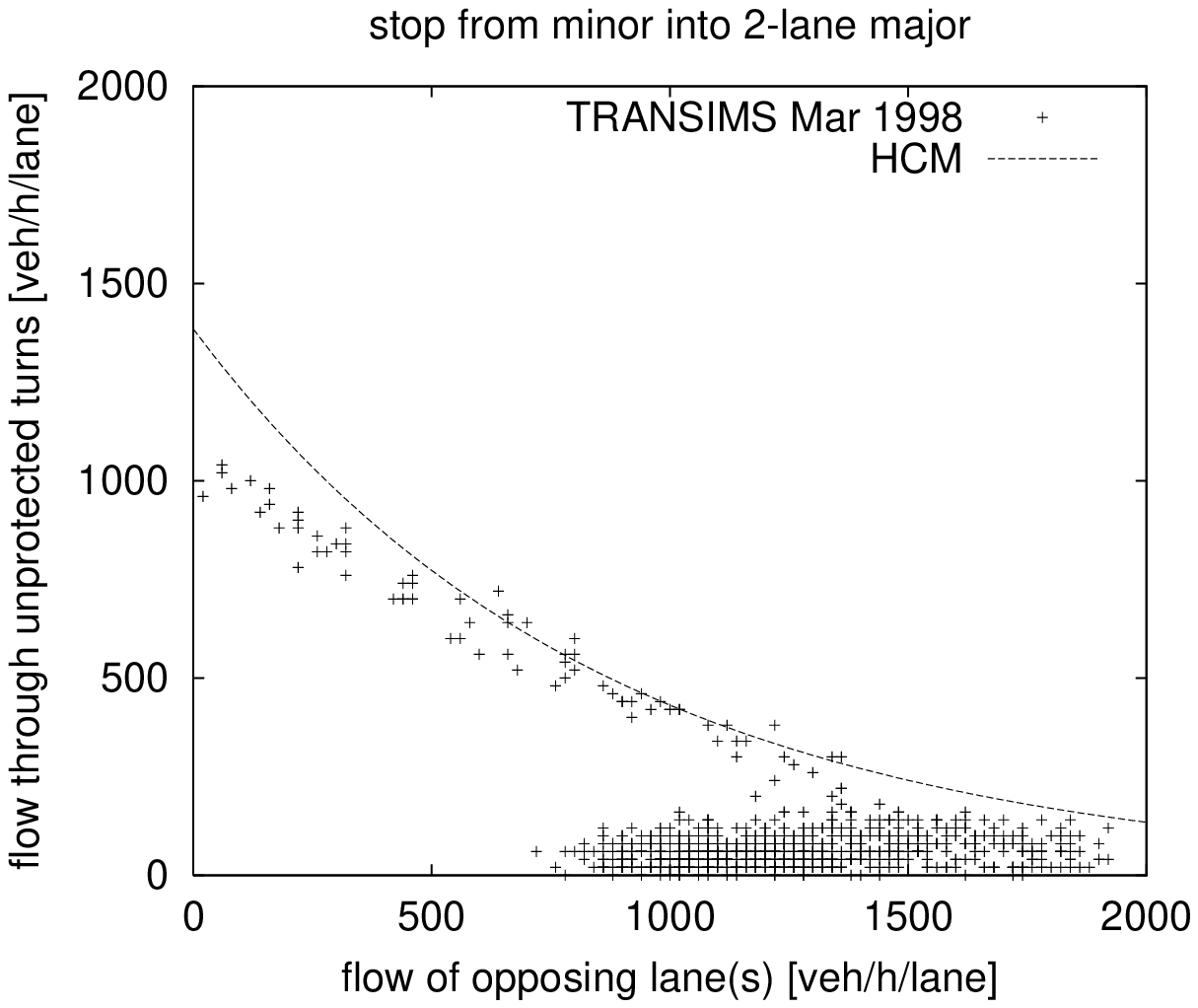}
}
\hfill
\parbox[c]{0.49\textwidth}{%
\centerline{%
\hfill
\includegraphics[width=0.42\textwidth]{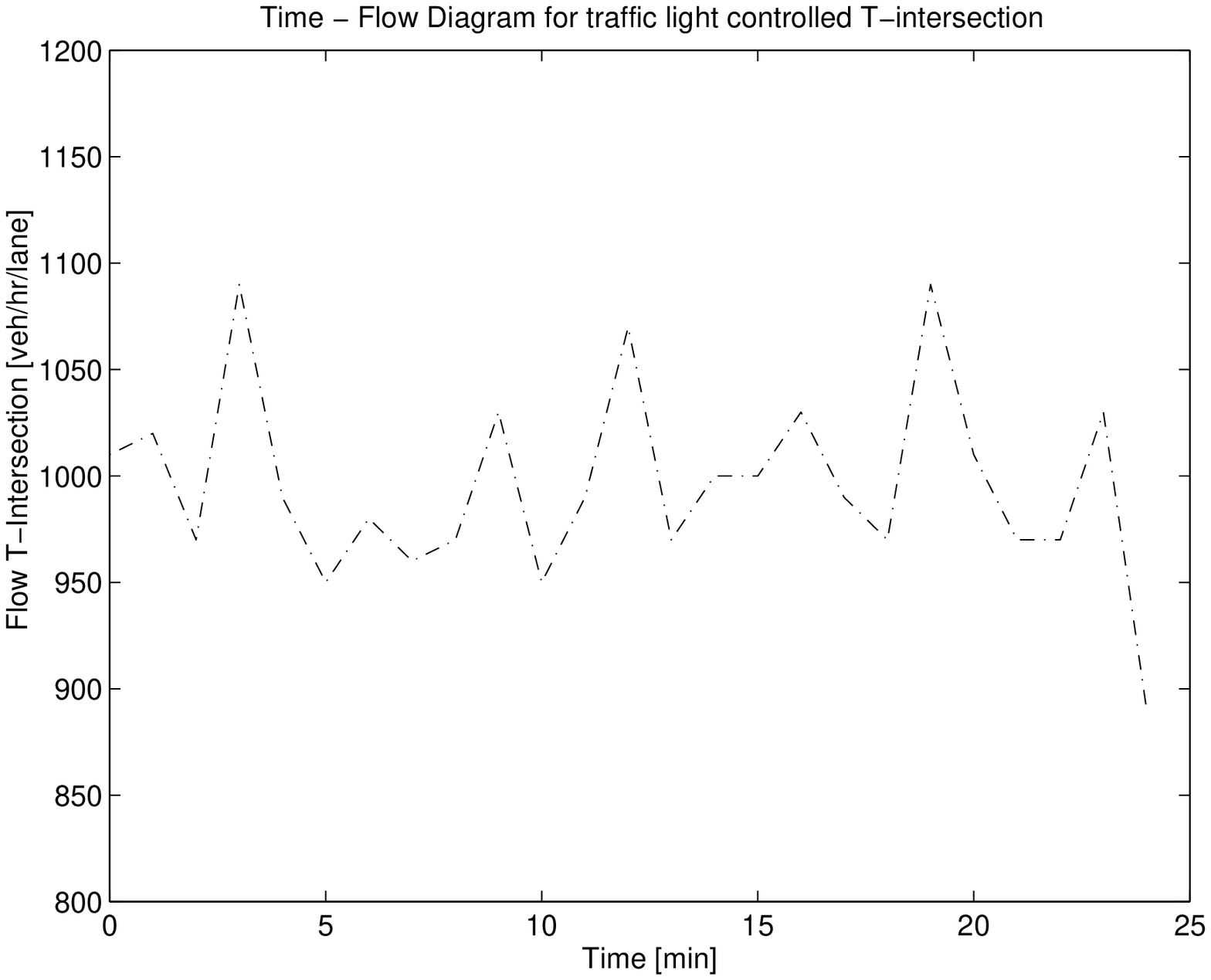}
\hspace{1pt}
}
}
}
    \caption{TRANSIMS macroscopic (emergent) traffic flow
      characteristics.  (a)~1-lane freeway.  (b)~3-lane freeway.
      (c)~Flow through stop sign onto 2-lane roadway. (d)~Flow through 
      traffic signal that is 30~sec red and 30~sec green, scaled to
      hourly flow rates.}
    \label{fig:flow-char}
\end{figure}

We show examples for the four categories (Fig.~\ref{fig:flow-char}):
(i)~Traffic in a 1-lane circle, thus validating the traffic flow
behavior of the car following implementation.  (ii)~Results of traffic
in a 3-lane circle, thus validating the addition of lane changing.
(iii)~Merge flows through a stop sign, thus validating the addition of
gap acceptance at unprotected turns.  (iv)~Flows through a traffic
light where vehicles need to be in the correct lanes for their
intended turns -- it thus simultaneously validates ``lane changing for
plan following'' and traffic light logic.

In our view, our validation results are within the range of field
measurements that one finds in the literature.  When going to a
specific study area, and depending on the specific question, more
calibration may become necessary, or in some cases additions to the
driving logic may be necessary.  For more information,
see~\cite{Nagel:etc:flow-char}.

\section{Graph partitioning}
\label{sec:partitioning}

Once we are able to handle split links, we need to partition the whole
transportation network graph in an efficient way.  Efficient means
several competing things: Minimize the number of split links; minimize
the number of other domains each CPU shares links with; equilibrate
the computational load as much as possible.

\begin{figure}[t]
  \begin{center}
    \includegraphics[width=\hsize]{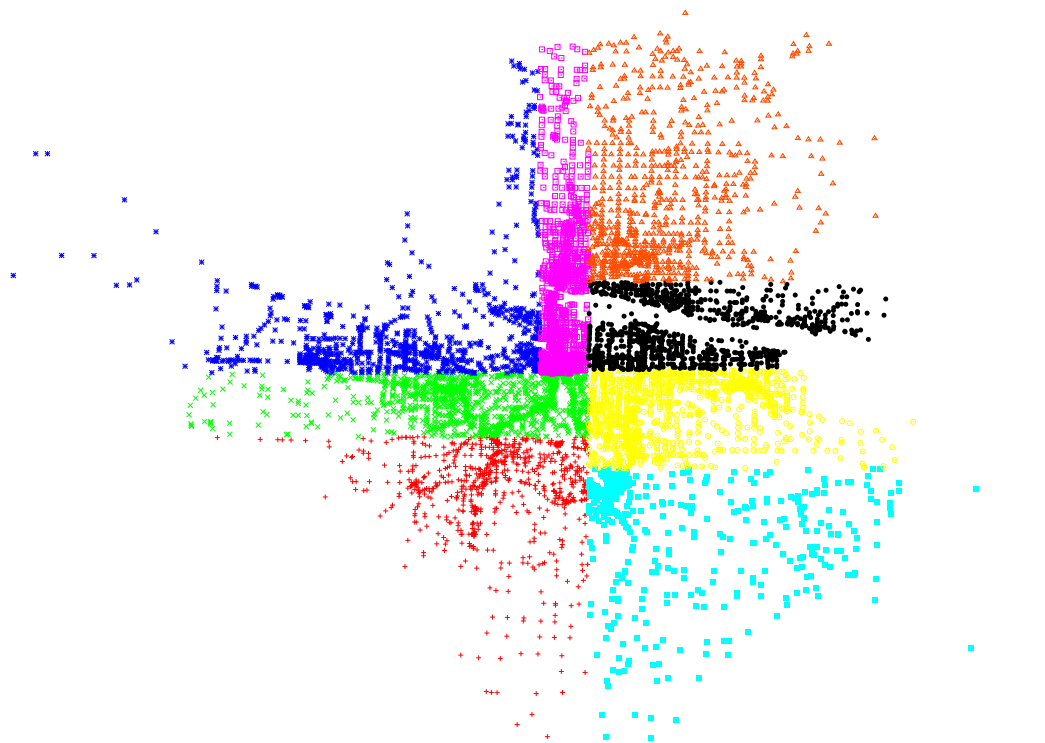}
    \caption{Orthogonal bi-section for Portland
      \portlandEUniLinks~links network.}
    \label{fig:ob-plot}
  \end{center}
\end{figure}

One approach to domain decomposition is orthogonal recursive
bi-section.  Although less efficient than METIS (explained below),
orthogonal bi-section is useful for explaining the general approach.
In our case, since we cut in the middle of links, the first step is to
accumulate computational loads at the nodes: each node gets a weight
corresponding to the computational load of all of its attached
half-links.  Nodes are located at their geographical coordinates.
Then, a vertical straight line is searched so that, as much as
possible, half of the computational load is on its right and the other
half on its left.  Then the larger of the two pieces is picked and cut
again, this time by a horizontal line.  This is recursively done until
as many domains are obtained as there are CPUs available, see
Fig.~\ref{fig:ob-plot}.  It is immediately clear that under normal
circumstances this will be most efficient for a number of CPUs that is
a power of two.  With orthogonal bi-section, we obtain compact and
localized domains, and the number of neighbor domains is limited.

Another option is to use the METIS library for graph partitioning (see
\cite{METIS} and references therein).  METIS uses multilevel
partitioning.  What that means is that first the graph is coarsened,
then the coarsened graph is partitioned, and then it is uncoarsened
again, while using an exchange heuristic at every uncoarsening step.
The coarsening can for example be done via random matching, which
means that first edges are randomly selected so that no two selected
links share the same vertex, and then the two nodes at the end of each
edge are collapsed into one.  Once the graph is sufficiently
collapsed, it is easy to find a good or optimal partitioning for the
collapsed graph.  During uncoarsening, it is systematically tried if
exchanges of nodes at the boundaries lead to improvements.
``Standard'' METIS uses multilevel recursive bisection: The initial
graph is partitioned into two pieces, each of the two pieces is
partitioned into two pieces each again, etc., until there are enough
pieces.  Each such split uses its own coarsening/uncoarsening
sequence.  $k$-METIS means that all $k$ partitions are found during a
single coarsening/uncoarsening sequence, which is considerably faster.
It also produces more consistent and better results for large $k$.

\begin{figure}[t]
  \begin{center}
    \includegraphics[width=\textwidth]{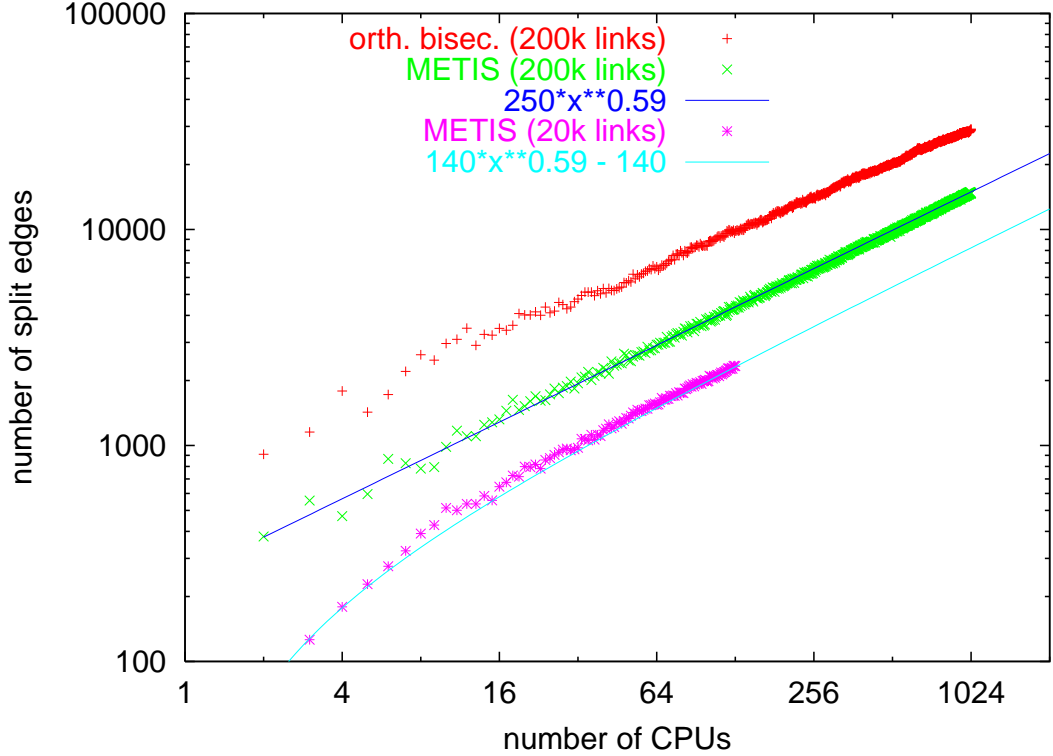}
    \caption{Number of split links as a function of the number of
      CPUs.  The top curve shows the result of orthogonal bisection
      for the \portlandAllUniLinks~links network.  The middle curve
      shows the result of METIS for the same network -- clearly, the
      use of METIS results in considerably fewer split links.  The
      bottom curve shows the result for the Portland
      \portlandEUniLinks~links network when again using METIS.  The
      theoretical scaling for orthogonal bisection is $N_{spl} \sim
      \sqrt{p}$, where $p$ is the number of CPUs.  Note that for $p
      \to N_{links}$, $N_{spl}$ needs to be the same for both graph
      partitioning methods.}
    \label{fig:splitedges}
  \end{center}
\end{figure}

\begin{figure}[t]
  \begin{center}
    \includegraphics[width=\hsize]{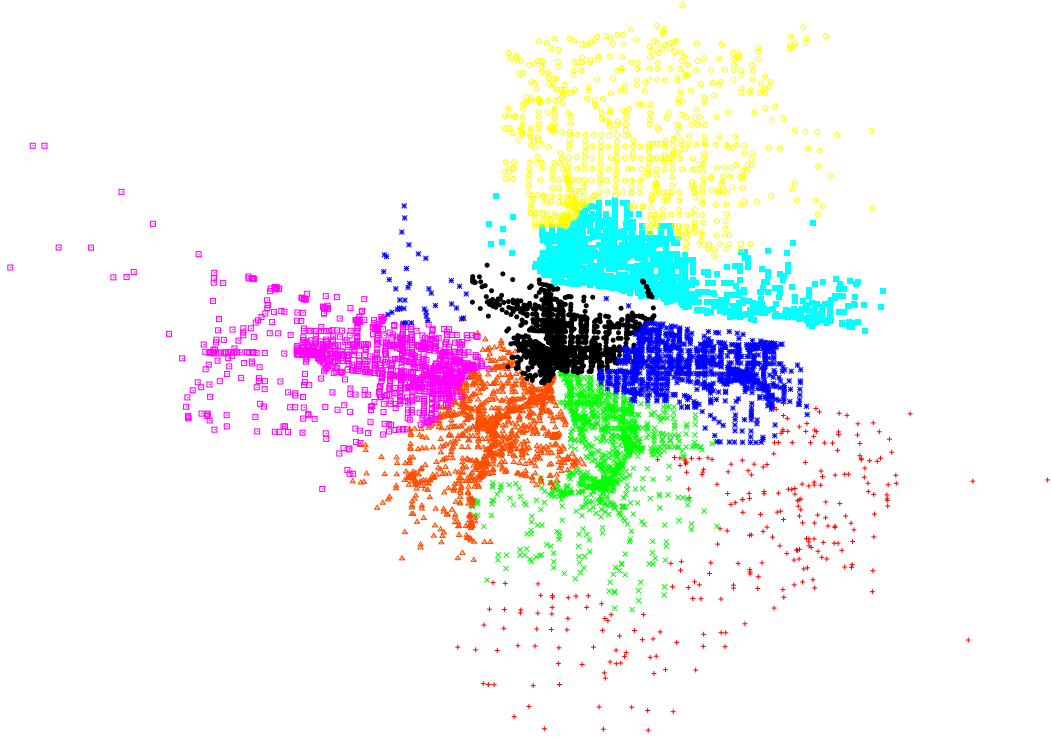}
    \caption{Partitioning by METIS.  Compare to
      Fig.~\protect\ref{fig:ob-plot}.}
    \label{fig:metis-plot}
  \end{center}
\end{figure}

METIS considerably reduces the number of split links, $N_{spl}$, as
shown in Fig.~\ref{fig:splitedges}.  The figure shows the number of
split links as a function of the number of domains for (i)~orthogonal
bi-section for a Portland network with \portlandAllUniLinks~links,
(ii)~METIS decomposition for the same network, and (iii)~METIS
decomposition for a Portland network with \portlandEUniLinks~links.
The network with \portlandAllUniLinks~links is derived from the TIGER
census data base, and will be used for the Portland case study for
TRANSIMS.  The network with \portlandEUniLinks~links is derived from
the EMME/2 network that Portland is currently using.
An example of the domains generated by METIS can be seen in
Fig.~\ref{fig:metis-plot}; for example, the algorithm now picks up the
fact that cutting along the rivers in Portland should be of advantage
since this results in a small number of split links.

We also show data fits to the METIS curves, $N_{spl} = 250 \,
p^{0.59}$ for the \portlandAllUniLinks~links network and $N_{spl} =
140 \, p^{0.59} - 140$ for the \portlandEUniLinks~links network, where
$p$ is the number of domains.  We are not aware of any theoretical
argument for the shapes of these curves for METIS.  It is however easy
to see that, for orthogonal bisection, the scaling of $N_{spl}$ has to
be $\sim p^{0.5}$.  Also, the limiting case where each node is on a
different CPU needs to have the same $N_{spl}$ both for bisection and
for METIS.  In consequence, it is plausible to use a scaling form of
$p^{\alpha}$ with $\alpha > 0.5$.  This is confirmed by the straight
line for large $p$ in the log-log-plot of Fig.~\ref{fig:splitedges}.
Since for $p=1$, the number of split links $N_{spl}$ should be zero,
for the \portlandEUniLinks~links network we use the equation $A \,
p^{\alpha} - A$, resulting in $N_{spl} = 140 \, p^{0.59}-140$ .  For
the \portlandAllUniLinks~links network, the resulting fit is so bad
that we did not add the negative term.  This leads to a kink for the
corresponding curves in Fig.~\ref{fig:other-rtr}.

\begin{figure}[t]
  \begin{center}
    \includegraphics[width=\textwidth]{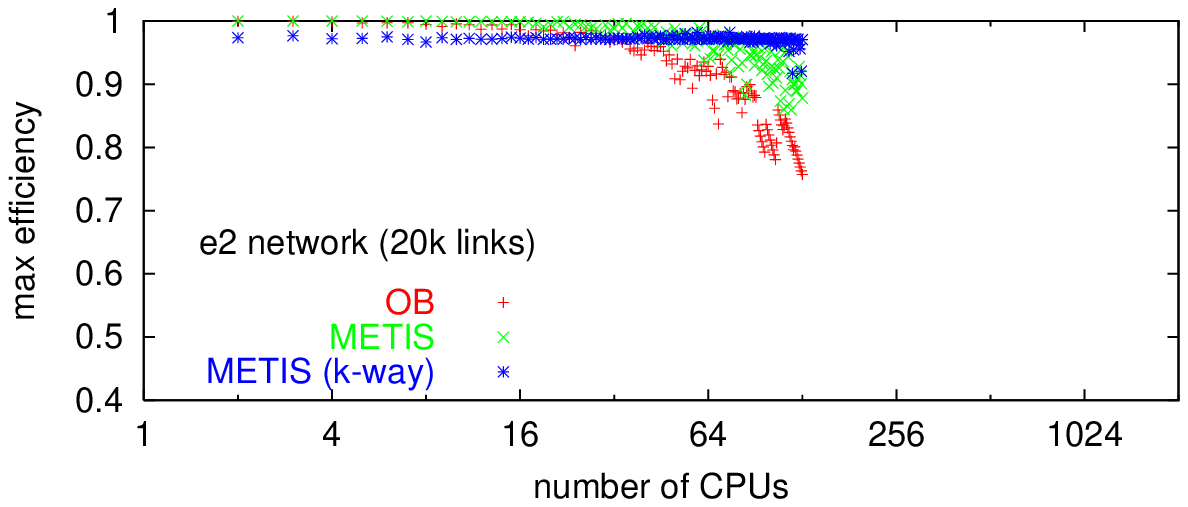}
    \includegraphics[width=\textwidth]{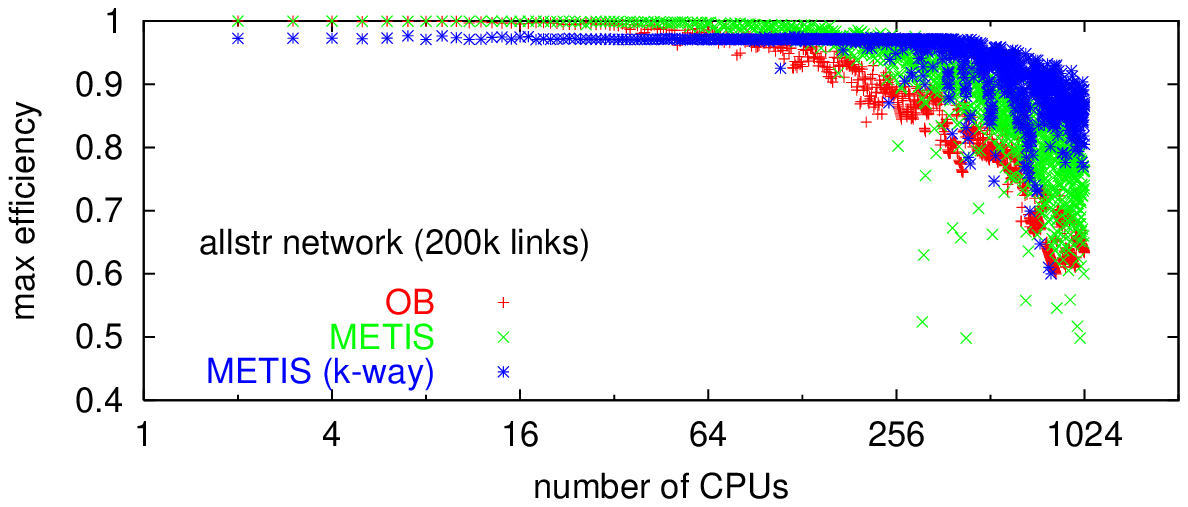}
    \caption{%
      \emph{Top:} Theoretical efficiency for Portland network with
      \portlandEUniLinks~links.  \emph{Bottom:} Theoretical efficiency
      for Portland network with \portlandAllUniLinks~links.  ``OB''
      refers to orthogonal bisection.  ``METIS (k-way)'' refers to an
      option in the METIS library.
}
    \label{fig:effic}
  \end{center}
\end{figure}

Such an investigation also allows to compute the theoretical
efficiency based on the graph partitioning.  Efficiency is optimal if
each CPU gets exactly the same computational load.  However, because
of the granularity of the entities (nodes plus attached half-links)
that we distribute, load imbalances are unavoidable, and they become
larger with more CPUs.  We define the resulting theoretical efficiency
due to the graph partitioning as
\begin{equation}
\label{eq:imbalance}
e_{dmn} := {\hbox{load on optimal partition} \over \hbox{load on largest
    partition}} \ ,
\end{equation}
where the load on the optimal partition is just the total load divided
by the number of CPUs.  We then calculated this number for actual
partitionings of both of our \portlandEUniLinks~links and of our
\portlandAllUniLinks~links Portland networks, see
Fig.~\ref{fig:effic}.  The result means that, according to this
measure alone, our \portlandEUniLinks~links network would still run
efficiently on 128~CPUs, and our \portlandAllUniLinks~links network
would run efficiently on up to 1024~CPUs.



\section{Adaptive Load Balancing}
\label{load_feedback}

\begin{figure}[t]
  \begin{center}
    \includegraphics[width=\hsize]{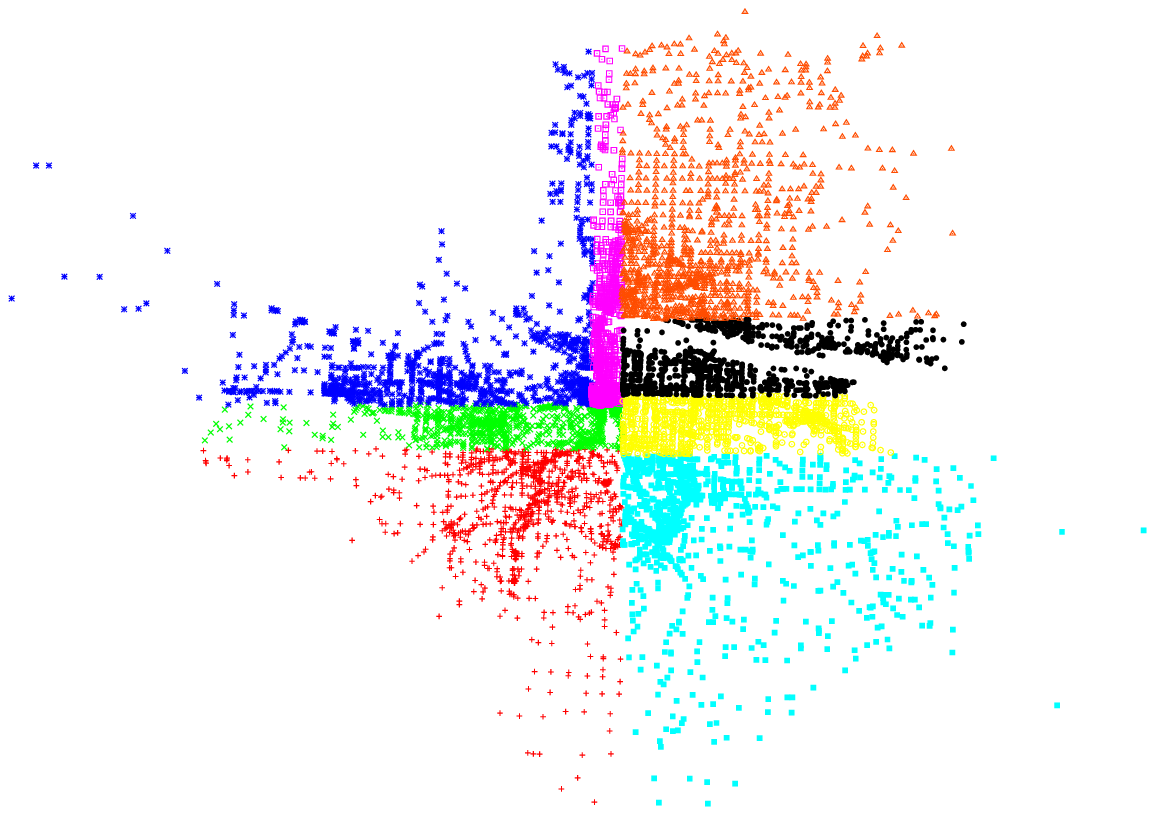}
    \caption{Partitioning after adaptive load balancing.  Compare to
      Fig.~\protect\ref{fig:ob-plot}. }
    \label{fig:ob-fb-plot}
  \end{center}
\end{figure}

In the last section, we explained how the street network is
partitioned into domains that can be loaded onto different CPUs.  In
order to be efficient, the loads on different CPUs should be as
similar as possible.  These loads do however depend on the actual
vehicle traffic in the respective domains.  Since we are doing
iterations, we are running similar traffic scenarios over and over
again.  We use this feature for an adaptive load balancing: During run
time we collect the execution time of each link and each intersection
(node). The statistics are output to file. For the next run of the
micro-simulation, the file is fed back to the partitioning algorithm.
In that iteration, instead of using the link lengths as load estimate,
the actual execution times are used as distribution criterion.
Fig.~\ref{fig:ob-fb-plot} shows the new domains after such a feedback
(compare to Fig.~\ref{fig:ob-plot}).

\begin{figure}
\centerline{%
\includegraphics[width=\hsize]{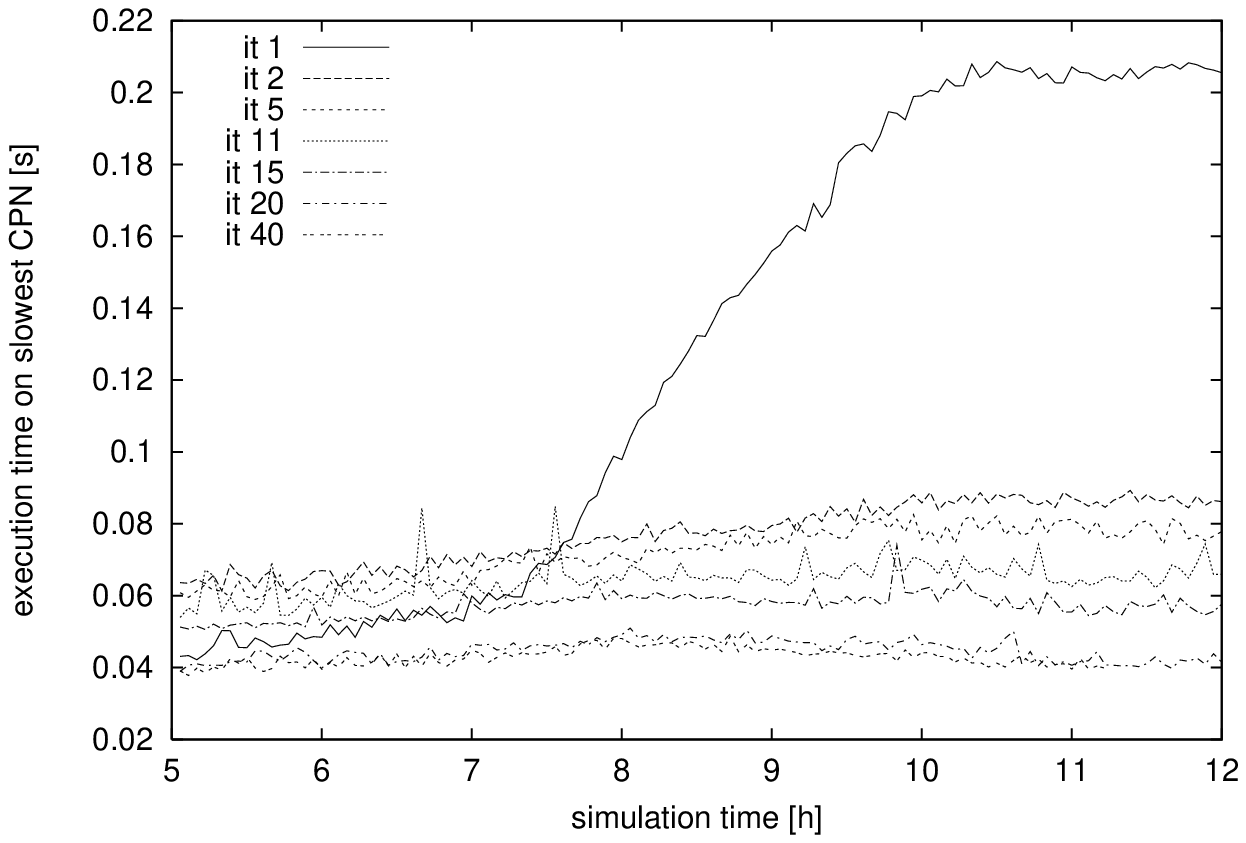}
}
\caption{\label{fig_load_feedback}%
Execution times with external load feedback.  These results were
obtained during the Dallas case
study~\protect\cite{Beckman:etc:case-study,Rickert:thesis}. 
}
\end{figure}

To verify the impact of this approach we monitored the execution times
per time-step throughout the simulation period.
Figure~\ref{fig_load_feedback} depicts the results of one of the
iteration series. For iteration~1, the load balancer uses the link
lengths as criterion. The execution times are low until congestion
appears around 7:30~am.  Then, the execution times increase fivefold
from 0.04~sec to 0.2~sec. In iteration~2 the execution times are
almost independent of the simulation time. Note that due to the
equilibration, the execution times for early simulation hours increase
from 0.04~sec to 0.06~sec, but this effect is more than compensated
later on.

The figure also contains plots for later iterations (11, 15, 20, and
40). The improvement of execution times is mainly due to the route
adaptation process: congestion is reduced and the average vehicle
density is lower.  On the machine sizes where we have tried it (up to
16~CPUs), adaptive load balancing led to performance improvements up
to a factor of 1.8.  It should become more important for larger
numbers of CPUs since load imbalances have a stronger effect there.


\section{Performance prediction for the TRANSIMS micro-simulation}
\label{sec:predict}

It is possible to systematically predict the performance of parallel
micro-si\-mu\-lations (e.g.~\cite{Jakobs:Gerling,Nagel:Schleicher}).
For this, several assumptions about the computer architecture need to
be made.  In the following, we demonstrate the derivation of such
predictive equations for coupled workstations and for parallel
supercomputers.

The method for this is to systematically calculate the wall clock time
for one time step of the micro-simulation.  We start by assuming that
the time for one time step has contributions from computation,
$T_{cmp}$, and from communication, $T_{cmm}$.  If these do not
overlap, as is reasonable to assume for coupled workstations, we have
\begin{equation}
T(p) = T_{cmp}(p) + T_{cmm}(p) \ ,
\end{equation}
where $p$ is the number of CPUs.\footnote{%
  For simplicity, we do not differentiate between CPUs and
  computational nodes.  Computational nodes can have more than one CPU 
  --- an example is a network of coupled PCs where each PC has Dual
  CPUs. 
}

Time for computation is assumed to follow
\begin{equation}
\label{eq:tcmp}
T_{cmp}(p) = {T_1 \over p} \cdot \Big( 1 + f_{ovr}(p) + f_{dmn}(p) \Big) \ .
\end{equation}
Here, $T_1$ is the time of the same code on one CPU (assuming a
problem size that fits on available computer memory); $p$ is the
number of CPUs; $f_{ovr}$ includes overhead effects (for example,
split links need to be administered by \emph{both} CPUs); $f_{dmn} =
1/e_{dmn} - 1$ includes the effect of unequal domain sizes discussed
in Sec.~\ref{sec:partitioning}.

Time for communication typically has two contributions: Latency and
bandwidth.  Latency is the time necessary to initiate the
communication, and in consequence it is independent of the message
size.  Bandwidth describes the number of bytes that can be
communicated per second.  So the time for one message is
\[
T_{msg} = T_{lt} + {S_{msg} \over b} \ ,
\]
where $T_{lt}$ is the latency, $S_{msg}$, is the message size, and 
$b$ is the bandwidth. 

However, for many of today's computer architectures, bandwidth is given
by at least two contributions: node bandwidth, and network bandwidth.
Node bandwidth is the bandwidth of the connection from the CPU to the
network.  If two computers communicate with each other, this is the
maximum bandwidth they can reach.  For that reason, this is sometimes
also called the ``point-to-point'' bandwidth.

The network bandwidth is given by the technology and topology of the
network.  Typical technologies are 10~Mbit Ethernet, 100~Mbit
Ethernet, FDDI, etc.  Typical topologies are bus topologies, switched
topologies, two-dimensional topologies (e.g.\ grid/torus), hypercube
topologies, etc.  A traditional Local Area Network (LAN) uses 10~Mbit
Ethernet, and it has a shared bus topology.  In a shared bus topology,
all communication goes over the same medium; that is, if several pairs
of computers communicate with each other, they have to share the
bandwidth.

For example, in our 100~Mbit FDDI network (i.e.\ a network bandwidth
of $b_{net} = 100$~Mbit) at Los Alamos National Laboratory, we found
node bandwidths of about $b_{nd} = 40$~Mbit.  That means that two
pairs of computers could communicate at full node bandwidth, i.e.\ 
using 80~of the 100~Mbit/sec, while three or more pairs were limited
by the network bandwidth.  For example, five pairs of computers could
maximally get $100/5 = 20$~Mbit/sec each.

A switched topology is similar to a bus topology, except that the
network bandwidth is given by the backplane of the switch.  Often, the
backplane bandwidth is high enough to have all nodes communicate with
each other at full node bandwidth, and for practical purposes one can
thus neglect the network bandwidth effect for switched networks.

If computers become massively parallel, switches with enough backplane
bandwidth become too expensive.  As a compromise, such supercomputers
usually use a communications topology where communication to
``nearby'' nodes can be done at full node bandwidth, whereas global
communication suffers some performance degradation.  Since we
partition our traffic simulations in a way that communication is
local, we can assume that we do communication with full node bandwidth
on a supercomputer.  That is, on a parallel supercomputer, we can
neglect the contribution coming from the $b_{net}$-term.  This
assumes, however, that the allocation of street network partitions to
computational nodes is done in some intelligent way which maintains
locality.

As a result of this discussion, we assume that the communication 
time per time step is
\[
T_{cmm}(p) = N_{sub} \cdot \Big( n_{nb}(p) \, T_{lt}
+ {N_{spl}(p) \over p} { S_{bnd} \over b_{nd} }
+ N_{spl}(p) \, { S_{bnd} \over b_{net} }\Big)\ ,
\]
which will be explained in the following paragraphs.  $N_{sub}$ is the
number of sub-time-steps. As discussed in Sec.~\ref{parallelization},
we do two boundary exchanges per time step, thus $N_{sub}=2$ for the
1999 TRANSIMS micro-simulation implementation.

$n_{nb}$ is the number of neighbor domains each CPU talks to.  All
information which goes to the same CPU is collected and sent as a
single message, thus incurring the latency only once per neighbor
domain.  For $p=1$, $n_{nb}$ is zero since there is no other domain to
communicate with.  For $p=2$, it is one.  For $p \to \infty$ and
assuming that domains are always connected, Euler's theorem for planar
graphs says that the average number of neighbors cannot become more
than six.  Based on a simple geometric argument, we use
\[
n_{nb}(p) = 2 \, (3 \sqrt{p} - 1) \, (\sqrt{p} - 1) / p \ ,
\]
which correctly has $n_{nb}(1) = 0$ and $n_{nb} \to 6$ for $p \to
\infty$.  Note that the METIS library for graph partitioning
(Sec.~\ref{sec:partitioning}) does not necessarily generate connected
partitions, making this potentially more complicated.


$T_{lt}$ is the latency (or start-up time) of each message.  $T_{lt}$
between 0.5 and 2~milliseconds are typical values for PVM on a
LAN~\cite{Rickert:thesis,Dongarra:etc:book}.

Next are the terms that describe our two bandwidth effects.
$N_{spl}(p)$ is the number of split links in the whole simulation;
this was already discussed in Sec.~\ref{sec:partitioning} (see
Fig.~\ref{fig:splitedges}).  Accordingly, $N_{spl}(p)/p$ is the number
of split links per computational node.  $S_{bnd}$ is the size of
the message per split link.  $b_{nd}$ and $b_{net}$ are the
node and network bandwidths, as discussed above.

In consequence, the combined time for one time step is
\[
T(p) = {T_1 \over p} \Big( 
1 + f_{ovr}(p) + f_{dmn}(p) \Big) +
\]
\[
N_{sub} \cdot \left(
n_{nb}(p) \,  T_{lt}
+ {N_{spl}(p) \over p} {S_{bnd} \over b_{nd}}
+ N_{spl}(p) \,  {S_{bnd} \over b_{net}} \right) \ .
\]

According to what we have discussed above, for $p \to \infty$ the
number of neighbors scales as $n_{nb} \sim const$ and the number of
split links in the simulation scales as $N_{spl} \sim
\sqrt{p}$.  In consequence for $f_{ovr}$ and $f_{dmn}$ small enough,
we have:
\begin{itemize}

\item for a shared or bus topology, $b_{net}$ is relatively small and
  constant, and thus
\[
T(p) \sim {1 \over p} + 1 + {1 \over \sqrt{p}} + \sqrt{p}
\to \sqrt{p} \ ;
\]

\item 
for a switched or a parallel supercomputer topology, we assume
$b_{net} = \infty$ and obtain
\[
T(p) \sim {1 \over p} + 1 + {1 \over \sqrt{p}} 
\to 1 \ .
\]
\end{itemize}
Thus, in a shared topology, adding CPUs will eventually
\emph{in}crease the simulation time, thus making the simulation
\emph{slower}.  In a non-shared topology, adding CPUs will eventually
not make the simulation any faster, but at least it will not be
detrimental to computational speed.  The dominant term in a shared
topology for $p \to \infty$ is the network bandwidth; the dominant
term in a non-shared topology is the latency.  

\begin{figure}[htbp]
  \begin{center}
    \includegraphics[width=0.8\hsize]{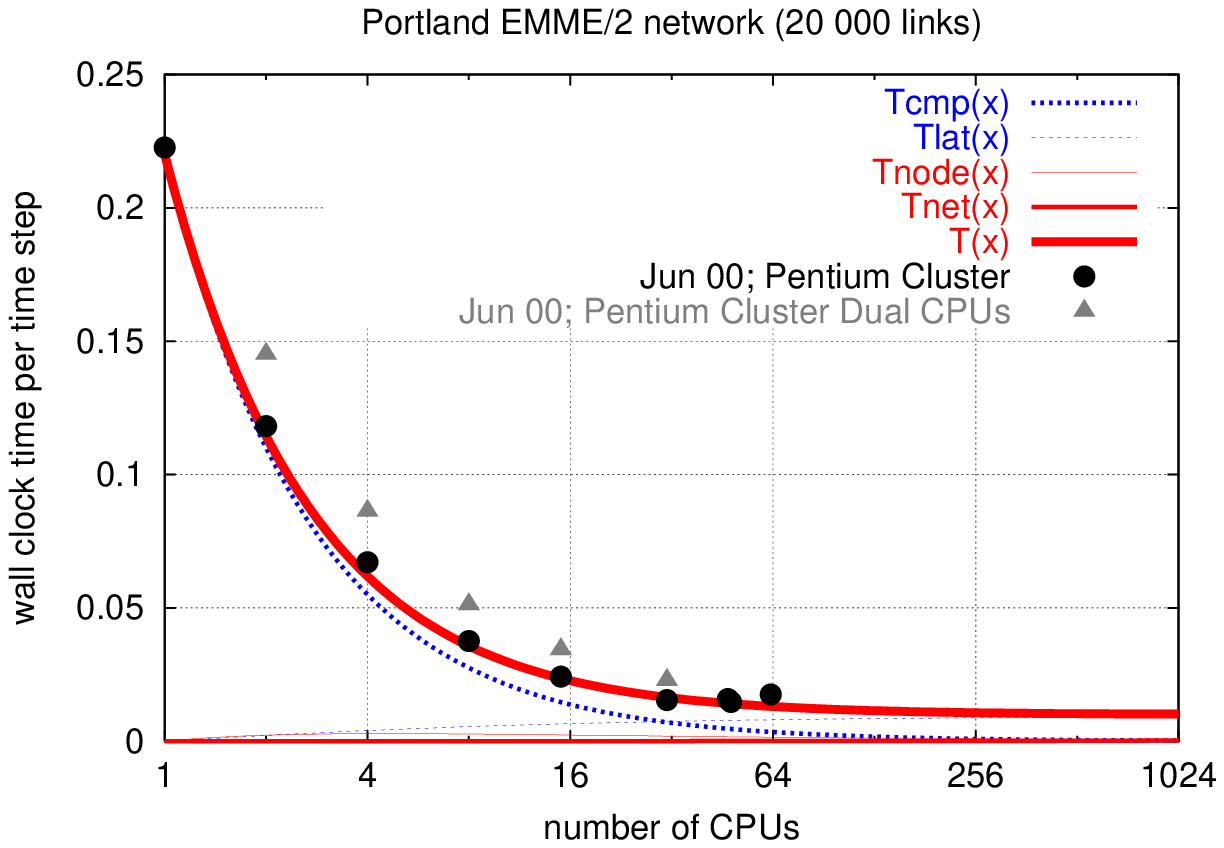}
    \includegraphics[width=0.8\hsize]{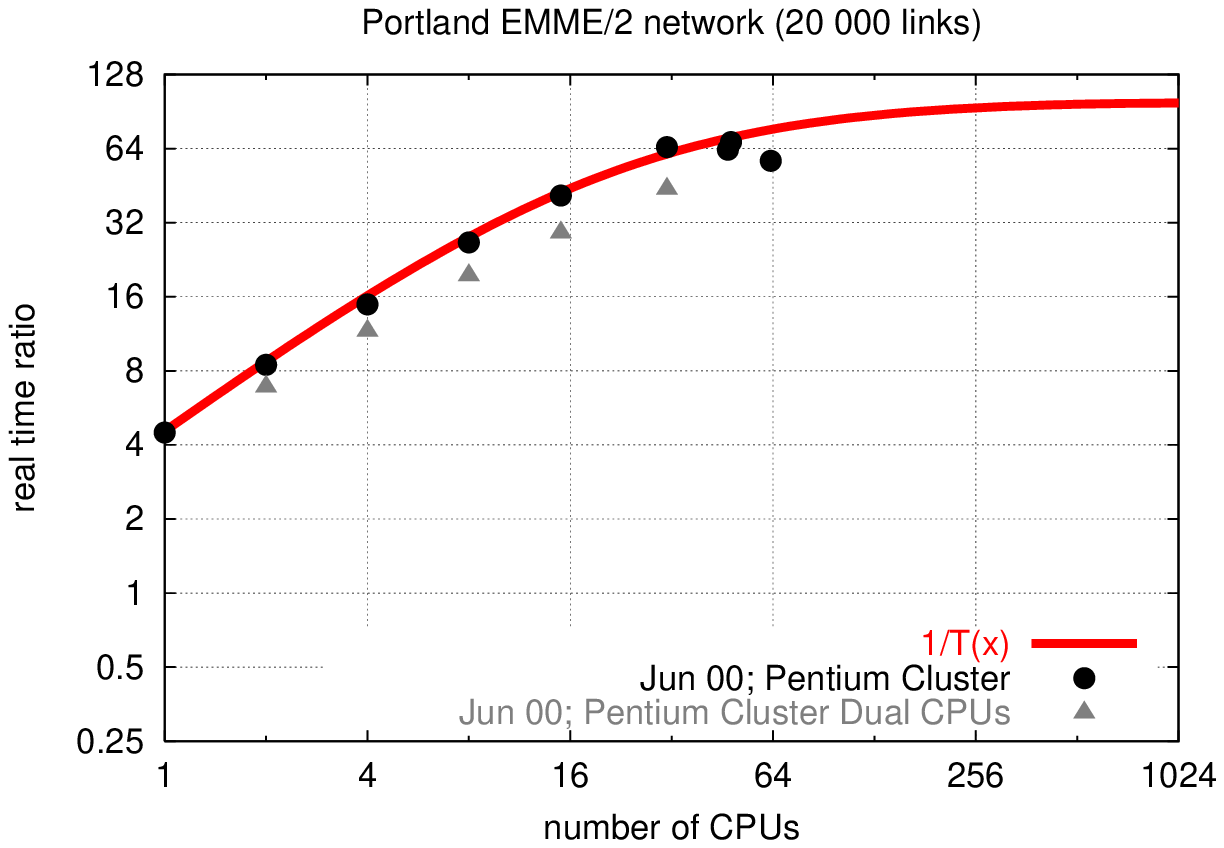}
    \caption{100~Mbit switched Ethernet LAN.  \emph{Top:} Individual time
      contributions.  \emph{Bottom:} Corresponding Real Time Ratios.
      The black dots refer to actually measured performance when using
      one CPU per cluster node; the crosses refer to actually measured
      performance when using dual CPUs per node (the $y$-axis still
      denotes the number of CPUs used).  The thick curve is the
      prediction according to the model.  The thin lines show the
      individual time contributions to the thick curve.
}
    \label{fig:100Mbit-switched}
  \end{center}
\end{figure}

The curves in Fig.~\ref{fig:100Mbit-switched} are results from this
prediction for a switched 100~Mbit Ethernet LAN; dots and crosses show
actual performance results.  The top graph shows the time for one time
step, i.e.\ $T(p)$, and the individual contributions to this value.
The bottom graph shows the real time ratio (RTR)
\[
rtr(p) := {\Delta t \over T(p)} = {1~sec \over T(p)} \ ,
\]
which says how much faster than reality the simulation is running.
$\Delta t$ is the duration a simulation time step, which is $1~sec$ in
TRANSIMS-1999.  The values of the free parameters are:\begin{itemize}
  
\item \textbf{Hardware-dependent parameters}.  We assume that the
  switch has enough bandwidth so that the effect of $b_{net}$ is
  negligeable.  Other hardware parameters are $T_{lt} = 0.8$~ms and
  $b_{nd} =
50$~Mbit/s.\footnote{%
  Our measurements have consistently shown that node bandwidths are
  lower than network bandwidths.  Even CISCO itself specifies
  148\,000~packets/sec, which translates to about 75~Mbit/sec, for the 
  100~Mbit switch that we use.
} 

\item \textbf{Implementation-dependent parameters}.  The number of message
  exchanges per time step is $N_{sub}=2$.
  
\item \textbf{Scenario-dependent parameters}.  Except when noted, our
  performance predictions and measurements refer to the Portland
  \portlandEUniLinks~links network.  We use, for the number of
  split links, $N_{spl}(p) = 140 \cdot p^{0.59} - 140$, as
  explained in Sec.~\ref{sec:partitioning}.
  
\item \textbf{Other Parameters.}  The message size depends on the
  plans format (which depends on the software design and
  implementation), on the typical number of links in a plan, and on
  the frequency per link of vehicles migrating from one CPU to
  another. We use $S_{bnd} = 200~Bytes$.  This is an average number;
  it includes all the information that needs to be sent when a vehicle
  migrates from one CPU to another.  The new TRANSIMS multi-modal
  plans format easily has 200~entries per driver and trip, resulting
  in 800~bytes of information just for the plan.  In addition, there
  is information about the vehicle (ID, speed, maximum acceleration,
  etc.); however, not in every time step a vehicle is migrated across
  a boundary on every split link.  In principle it is however possible
  to compress the plans information, so improvements are possible here
  in the future.  Also, we have not explicitely modelled simulation
  output, which is indeed a performance issue on Beowulf clusters.

\end{itemize}
These parameters were obtained in the following way: First, we
obtained plausible values via systematic communication tests using
messages similar to the ones used in the actual
simulation~\cite{Rickert:thesis}.  Then, we ran the simulation without
any vehicles (see below) and adapted our values accordingly.  Running
the simulation without vehicles means that we have a much better
control of $S_{bnd}$.  In practice, the main result of this step was
to set $t_{lat}$ to 0.8~msec, which is plausible when compared to the
hardware value of 0.5~msec.  Last, we ran the simulations with
vehicles and adjusted $S_{bnd}$ to fit the data. --- In consequence,
for the switched 100~Mbit Ethernet configurations, within the data
range our curves are model fits to the data.  Outside the data range
and for other configurations, the curves are model-based predictions.

The plot (Fig.~\ref{fig:100Mbit-switched}) shows that even something
as relatively profane as a combination of regular Pentium CPUs using a
switched 100Mbit Ethernet technology is quite capable in reaching good
computational speeds.  For example, with 16~CPUs the simulation runs
40~times faster than real time; the simulation of a 24~hour time
period would thus take 0.6~hours.  These numbers refer, as said above,
to the Portland \portlandEUniLinks~links network.  Included in the
plot (black dots) are measurements with a compute cluster that
corresponds to this architecture.  The triangles with lower
performance for the same number of CPUs come from using dual instead
of single CPUs on the computational nodes.  Note that the curve levels
out at about forty times faster than real time, no matter what the
number of CPUs.  As one can see in the top figure, the reason is the
latency term, which eventually consumes nearly all the time for a time
step.  This is one of the important elements where parallel
supercomputers are different: For example the Cray T3D has a more than
a factor of ten lower latency under PVM~\cite{Dongarra:etc:book}.

\begin{figure}[htbp]
  \begin{center}
    \includegraphics[width=0.8\hsize]{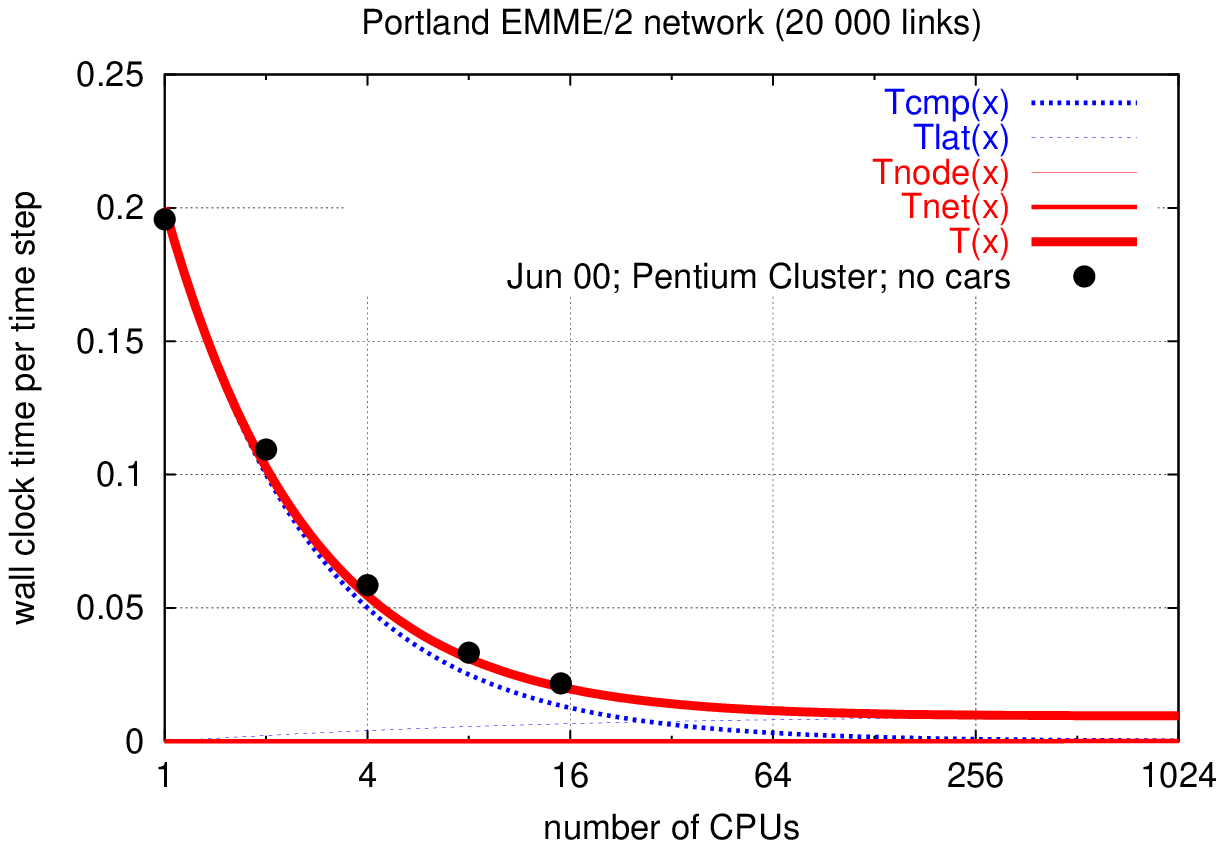}
    \includegraphics[width=0.8\hsize]{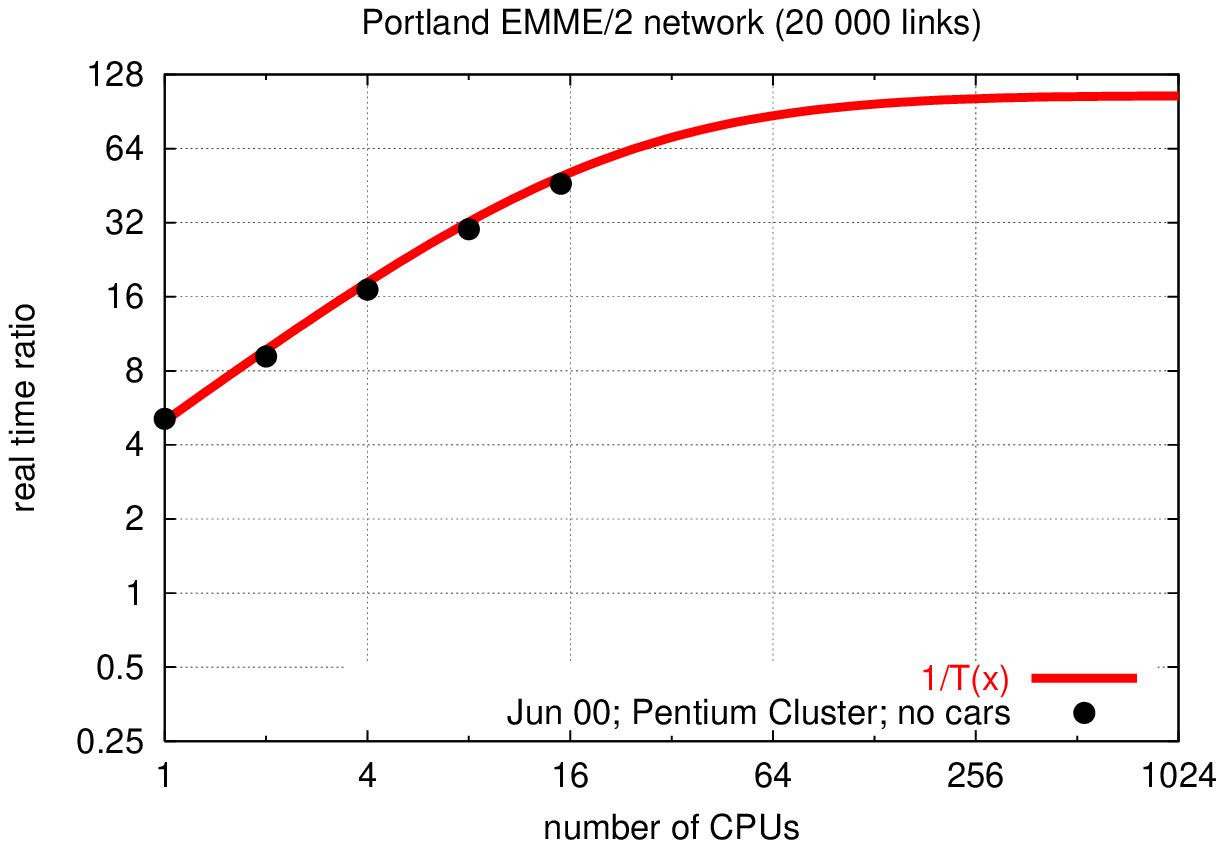}
    \caption{%
      100~Mbit switched Ethernet LAN; simulation without vehicles.
      \emph{Top:} Individual time contributions.  \emph{Bottom:}
      Corresponding Real Time Ratios.  The same remarks as to
      Fig.~\ref{fig:100Mbit-switched} apply.  In particular, black
      dots show measured performance, whereas curves show predicted
      performance.
}
    \label{fig:nocars}
  \end{center}
\end{figure}

As mentioned above, we also ran the same simulation without any
vehicles.  In the TRANSIMS-1999 implementation, the simulation sends
the contents of each CA boundary region to the neighboring CPU even
when the boundary region is empty.
%
%
Without compression, this is five integers for five sites, times the
number of lanes, resulting in about 40~bytes per split edge, which is
considerably less than the 800~bytes from above.  The results are
shown in Fig.~\ref{fig:nocars}.  Shown are the computing times with
1~to~15 single-CPU slaves, and the corresponding real time ratio.
Clearly, we reach better speed-up without vehicles than with vehicles
(compare to Fig.~\ref{fig:100Mbit-switched}).  Interestingly, this
does not matter for the maximum computational speed that can be
reached with this architecture: Both with and without vehicles, the
maximum real time ratio is about~80; it is simply reached with a
higher number of CPUs for the simulation with vehicles.  The reason is
that eventually the only limiting factor is the network latency term,
which does not have anything to do with the \emph{amount} of
information that is communicated.

\begin{figure}[htb]
  \begin{center}
    \includegraphics[width=0.8\hsize]{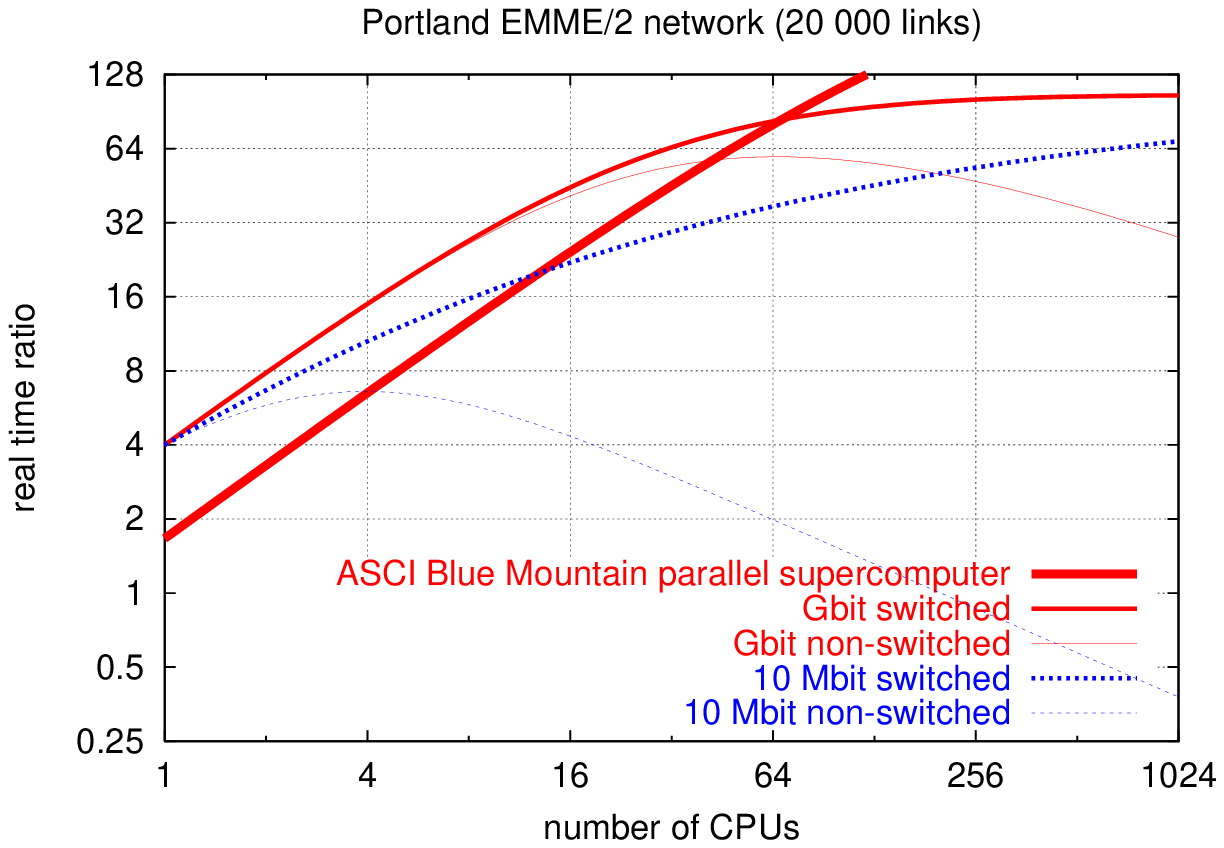}
    \includegraphics[width=0.8\hsize]{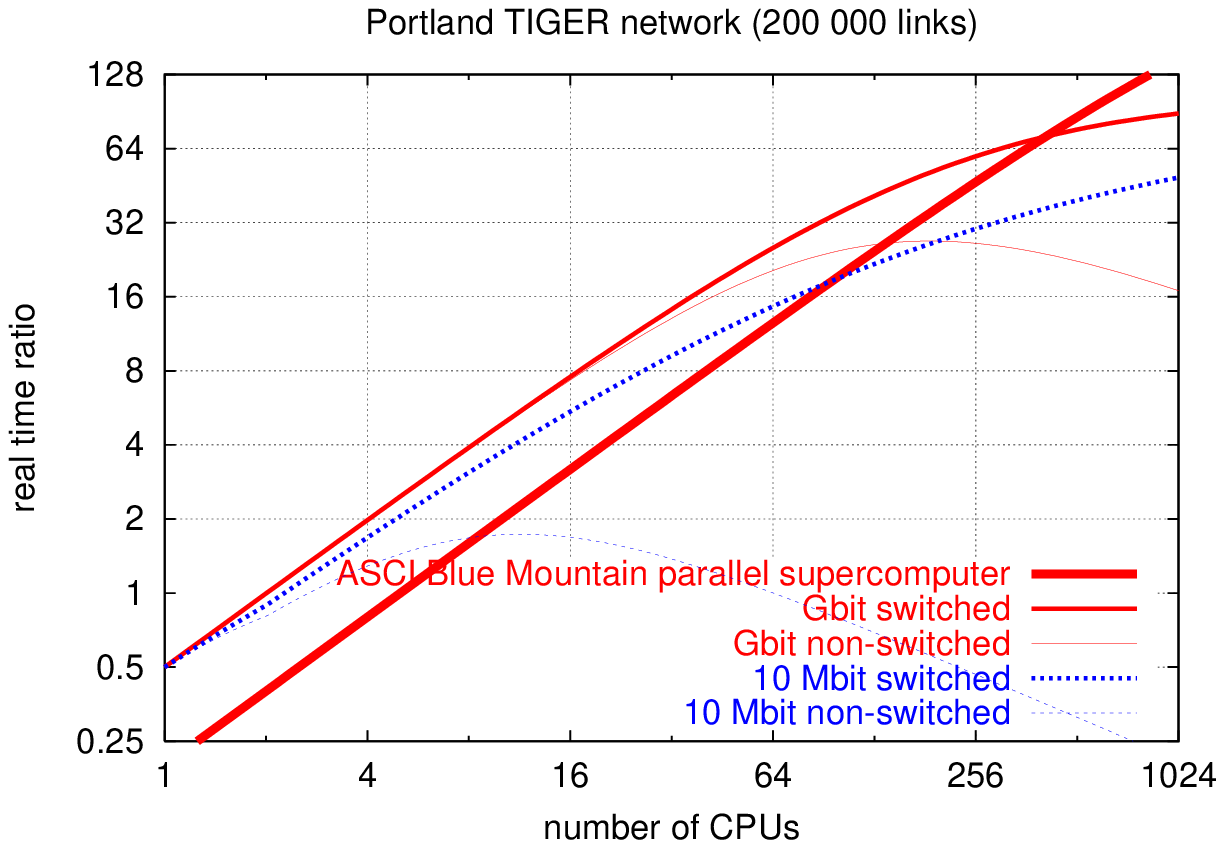}
    \caption{%
      Predictions of real time ratio for other computer
      configurations. \emph{Top:} With Portland EMME/2 network
      (\portlandEUniLinks~links).  \emph{Bottom:} With Portland TIGER
      network (\portlandAllUniLinks~links).  Note that for the
      switched configurations and for the supercomputer, the
      saturating real time ratio is the same for both network sizes,
      but it is reached with different numbers of CPUs.  This behavior
      is typical for parallel computers: They are particularly good at
      running larger and larger problems within the same computing
      time. --- All curves in both graphs are predictions from our
      model.  We have some performance measurements for the ASCI
      maschine, but since they were done with an older and slower
      version of the code, they are omitted in order to avoid
      confusion.
}
    \label{fig:other-rtr}
  \end{center}
\end{figure}

Fig.~\ref{fig:other-rtr}~(top) shows some predicted real time ratios
for other computing architectures.  For simplicity, we assume that all
of them except for one special case explained below use the same
500~MHz Pentium compute nodes.  The difference is in the networks: We
assume 10~Mbit non-switched, 10~Mbit switched, 1~Gbit non-switched,
and 1~Gbit switched.  The curves for 100~Mbit are in between and were
left out for clarity; values for switched 100~Mbit Ethernet were
already in Fig.~\ref{fig:100Mbit-switched}.  One clearly sees that for
this problem and with today's computers, it is nearly impossible to
reach \emph{any} speed-up on a 10~Mbit Ethernet, even when switched.
Gbit Ethernet is somewhat more efficient than 100~Mbit Ethernet for
small numbers of CPUs, but for larger numbers of CPUs, switched Gbit
Ethernet saturates at exactly the same computational speed as the
switched 100~Mbit Ethernet.  This is due to the fact that we assume
that latency remains the same -- after all, there was no improvement
in latency when moving from 10~to 100~Mbit Ethernet.  FDDI is
supposedly even worse~\cite{Dongarra:etc:book}.

The thick line in Fig.~\ref{fig:other-rtr} corresponds to the ASCI
Blue Mountain parallel supercomputer at Los Alamos National
Laboratory.  On a per-CPU basis, this machine is slower than a 500~MHz
Pentium.  The higher bandwidth and in particular the lower latency
make it possible to use higher numbers of CPUs efficiently, and in
fact one should be able to reach a real time ratio of 128 according to
this plot.  By then, however, the granularity effect of the unequal
domains (Eq.~(\ref{eq:imbalance}), Fig.~\ref{fig:effic}) would have
set in, limiting the computational speed probably to about 100~times
real time with 128~CPUs.  We actually have some speed measurements on
that machine for up to 96~CPUs, but with a considerably slower code
from summer~1998.   We omit those values from the plot in order to
avoid confusion. 

Fig.~\ref{fig:other-rtr}~(bottom) shows predictions for the higher
fidelity Portland \portlandAllUniLinks~links network with the same
computer architectures.  The assumption was that the time for one time
step, i.e.\ $T_1$ of Eq.~(\ref{eq:tcmp}), increases by a factor of
eight due to the increased load.  This has not been verified yet.
However, the general message does not depend on the particular
details: When problems become larger, then larger numbers of CPUs
become more efficient.  Note that we again saturate, with the switched
Ethernet architecture, at 80~times faster than real time, but this
time we need about 64~CPUs with switched Gbit Ethernet in order to get
40~times faster than real time --- for the smaller Portland
\portlandEUniLinks~links network with switched Gbit Ethernet we would
need 8~of the same CPUs to reach the same real time ratio.  In short
and somewhat simplified: As long as we have enough CPUs, we can
micro-simulate road networks of \emph{arbitrarily largesize}, with
hundreds of thousands of links and more, 40~times faster than real
time, even without supercomputer hardware.  --- Based on our
experience, we are confident that these predictions will be lower
bounds on performance: In the past, we have always found ways to make
the code more efficient.


\section{Speed-up and efficiency}
\label{sec:speedup}

We have cast our results in terms of the real time ratio, since this
is the most important quantity when one wants to get a practical study
done.  In this section, we will translate our results into numbers of
speed-up, efficiency, and scale-up, which allow easier comparison for
computing people.

Let us define speed-up as
\[
S(p) := {T(1) \over T(p)}  \ ,
\]
where $p$ is again the number of CPUs, $T(1)$ is the time for one
time-step on one CPU, and $T(p)$ is the time for one time step on $p$
CPUs.  Depending on the viewpoint, for $T(1)$ one uses either the
running time of the parallel algorithm on a single CPU, or the fastest
existing sequential algorithm.  Since TRANSIMS has been designed for
parallel computing and since there is no sequential simulation with
exactly the same properties, $T(1)$ will be the running time of the
parallel algorithm on a single CPU.  For time-stepped simulations such
as used here, the difference is expected to be small.\footnote{%
  An event-driven simulation could be a counter-example: Depending on
  the implementation, it could be extremely fast on a single CPU up to 
  medium problem sizes, but slow on a parallel machine.
}

Now note again that the real time ratio is 
$
rtr(p) = {1~sec / T(p)} \ .
$
Thus, in order to obtain the speed-up from the real time ratio, one
has to multiply all real time ratios by $T(1)/(1~sec)$.  On a
logarithmic scale, a multiplication corresponds to a linear shift.  In
consequence, speed-up curves can be obtained from our real time ratio
curves by shifting the curves up or down so that they start at one.

\begin{figure}[htbp]
  \begin{center}
    \includegraphics[width=0.8\hsize]{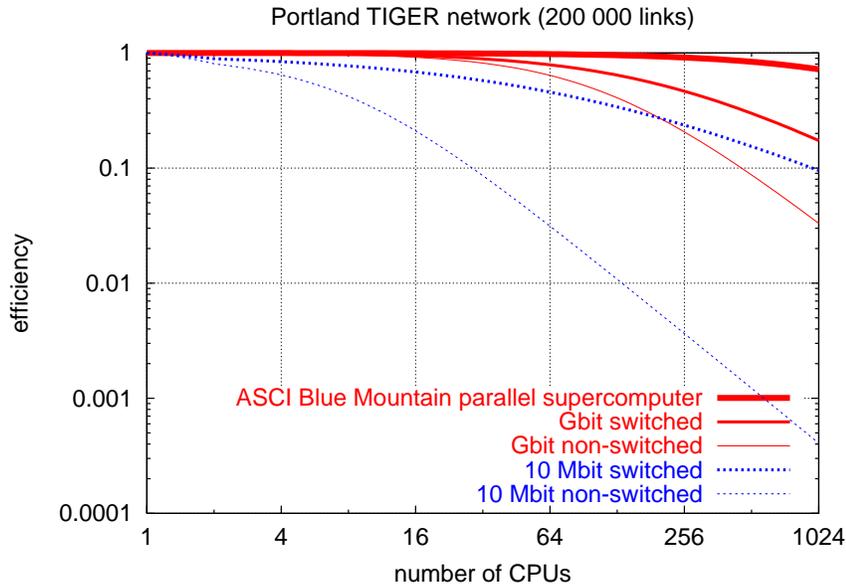}
    \caption{Efficiency for the same configurations as in Fig.~\ref{fig:other-rtr}~bottom.   Note that the curves contain exactly the same information.}
    \label{fig:eff}
  \end{center}
\end{figure}

This also makes it easy to judge if our speed-up is linear or not.
For example in Fig.~\ref{fig:other-rtr} bottom, the curve which starts 
at 0.5 for 1~CPU should have an RTR of 2 at 4~CPU, an RTR of 8 at
16~CPUs, etc.  Downward deviations from this mean sub-linear speed-up.
Such deviations are commonly described by another number, called
efficiency, and defined as
\[
E(p) := {T(1)/p \over T(p)} \ .
\]
Fig.~\ref{fig:eff} contains an example. Note that this number contains
no new information; it is just a re-interpretation.  Also note that in
our logarithmic plots, $E(p)$ will just be the difference to the
diagonal $p \, T(1)$.  Efficiency can point out where improvements
would be useful.

\section{Other modules}
\label{sec:other}

As explained in the introduction, a micro-simulation in a software
suite for transportation planning would have to be run many times
(``feedback iterations'') in order to achieve consistency between
modules.  For the microsimulation alone, and assuming our
16~CPU-machine with switched 100~Mbit Ethernet, we would need about
30~hours of computing time in order to simulate 24~hours of traffic
fifty times in a row.  In addition, we have the contributions from the
other modules (routing, activities generation).  In the past, these
have never been a larger problem than the micro-simulation, for
several reasons:\begin{itemize}

\item The algorithms of the other modules by themselves did
  significantly less computation than the micro-simulation.

\item Even when these algorithms start using considerable amounts of
  computer time, they are ``trivially'' parallelizable by simply
  distributing the households across CPUs.\footnote{%
    This is possible because of the specific purpose TRANSIMS is
    designed for.  In real time applications, where absolute speed
    between request and response matters, the situation is
    different~\cite{Chabini:recycle:trb}.
}

\item In addition, during the iterations we never replan more than
  about 10\% of the population, saving additional computer time.

\end{itemize}
In summary, the TRANSIMS modules besides the traffic micro-simulation
currently do not contribute significantly to the computational burden;
in consequence, the computational performance of the traffic
micro-simulation is a good indicator of the overall performance of the 
simulation system.

\section{Summary}

This paper explains the parallel implementation of the TRANSIMS
micro-simulation.  Since other modules are computationally less
demanding and also simpler to parallelize, the parallel implementation
of the micro-simulation is the most important and most complicated
piece of parallelization work.  The parallelization method for the
TRANSIMS micro-simulation is domain decomposition, that is, the
network graph is cut into as many domains as there are CPUs, and each
CPU simulates the traffic on its domain.  We cut the network graph in
the middle of the links rather than at nodes (intersections), in order
to separate the traffic dynamics complexity at intersections from the
complexity of the parallel implementation.  We explain how the
cellular automata (CA) or any technique with a similar time depencency
scheduling helps to design such split links, and how the message 
exchange in TRANSIMS works.

The network graph needs to be partitioned into domains in a way that
the time for message exchange is minimized.  TRANSIMS uses the METIS
library for this goal.  Based on partitionings of two different
networks of Portland (Oregon), we calculate the number of CPUs where
this approach would become inefficient just due to this criterion.
For a network with \portlandAllUniLinks~links, we find that due to
this criterion alone, up to 1024~CPUs would be efficient.  We also
explain how the TRANSIMS micro-simulation adapts the partitions from
one run to the next during feedback iterations (adaptive load
balancing).

We finally demonstrate how computing time for the TRANSIMS
micro-simulation (and therefore for all of TRANSIMS) can be
systematically predicted.  An important result is that the Portland
\portlandEUniLinks~links network runs about 40~times faster than real
time on 16~dual 500~MHz Pentium computers connected via switched
100~Mbit Ethernet. These are regular desktop/LAN technologies.  When
using the next generation of communications technology, i.e.\ Gbit
Ethernet, we predict the same computing speed for a much larger
network of \portlandAllUniLinks~links with 64~CPUs.

\section{Acknowledgments}

This is a continuation of work that was started at Los Alamos National
Laboratory (New Mexico) and at the University of Cologne (Germany).
An earlier version of some of the same material can be found in
Ref.~\cite{Rickert:Nagel:hpcn99}.  We thank the U.S.\ Federal
Department of Transportation and Los Alamos National Laboratory for
making TRANSIMS available free of charge to academic institutions.
The version used for this work was ``TRANSIMS-LANL Version 1.0''.


\bibliographystyle{parallel}
\bibliography{ref,kai,xref}

\end{document}